\documentclass[
a4paper,
onecolumn,
table,
superscriptaddress,
amsmath,
amssymb,
aps,
pra,
accepted=2021-01-20
]{quantumarticle}
\pdfoutput=1
\usepackage[utf8]{inputenc}
\usepackage[english]{babel}
\usepackage[T1]{fontenc}

\usepackage{mdframed}
\usepackage{booktabs}
\usepackage{longtable}

\usepackage{hyperref}

\usepackage[numbers]{natbib}
\usepackage[dvipsnames]{xcolor}
\usepackage{tikz}
\usetikzlibrary{automata, arrows}
\usetikzlibrary{decorations.pathreplacing,angles,quotes} %braces in tikz
\usetikzlibrary{positioning}
\usetikzlibrary{shapes,arrows}
\usetikzlibrary{decorations.shapes}
\usetikzlibrary{arrows.meta}
\usetikzlibrary{backgrounds, fit}
\usetikzlibrary{calc}
\tikzstyle{decision} = [diamond, draw, fill=blue!20, 
text width=4.5em, text badly centered, node distance=3cm, inner sep=0pt]
\tikzstyle{block} = [rectangle, draw, fill=blue!20, 
text width=5em, text centered, rounded corners, minimum height=4em]
\tikzstyle{line} = [draw, -latex']
\tikzstyle{cloud} = [draw, ellipse,fill=red!20, node distance=3cm,
minimum height=2em]

\usepackage{graphicx}
\usepackage[position=top, caption=false]{subfig}
\usepackage{enumitem}
\usepackage{algpseudocode}
\usepackage{algorithm}
\usepackage{amssymb}
\usepackage{mathtools}
\usepackage{bbold}

\usepackage{amsthm}
\usepackage{amsmath}
\usepackage{float}

\usepackage[hang,flushmargin]{footmisc}

\theoremstyle{definition}

\begin{document}

\title{Cost-optimal single-qubit gate synthesis in the Clifford hierarchy%\thanks{Grants or other notes
  % about the article that should go on the front page should be
  % placed here. General acknowledgments should be placed at the end of the article.}
}
% \titlerunning{Short form of title}        % if too long for running head

%\author{Gary J. Mooney$^{1,\, \dag}$, Charles D. Hill$^{1, \,2, \,\ddagger}$, and Lloyd C.L. Hollenberg$^{1, \,\S}$}

% Specify affiliations first to customise order.
%\affiliation{School of Physics, University of Melbourne, VIC, Parkville, 3010, Australia.}
%\affiliation{School of Mathematics and Statistics, University of Melbourne, VIC, Parkville, 3010, Australia.}

\author{Gary J. Mooney}
\email{mooneyg@unimelb.edu.au}
\orcid{0000-0002-3253-9815}
\affiliation{School of Physics, University of Melbourne, VIC, Parkville, 3010, Australia.}
\author{Charles D. Hill}
\email{cdhill@unimelb.edu.au}
\orcid{0000-0003-0185-8028}
\affiliation{School of Physics, University of Melbourne, VIC, Parkville, 3010, Australia.}
\affiliation{School of Mathematics and Statistics, University of Melbourne, VIC, Parkville, 3010, Australia.}
\author{Lloyd C.L. Hollenberg}
\email{lloydch@unimelb.edu.au}
\orcid{0000-0001-7672-6965}
\affiliation{School of Physics, University of Melbourne, VIC, Parkville, 3010, Australia.}

%\author{Charles D. Hill$^{1, \,2, \,\ddagger}$}
%\author[1, 3, $\S$]{Lloyd C.L. Hollenberg}

%\footnote{$^1$School of Physics, University of Melbourne, VIC, Parkville, 3010, Australia.\\ $^2$School of Mathematics and Statistics, University of Melbourne, VIC, Parkville, 3010, Australia.\\
%$^\dag$gmooney@student.unimelb.edu.au\\
%$^\ddagger$cdhill@unimelb.edu.au\\
%$^\S$lloydch@unimelb.edu.au}

% \authorrunning{Short form of author list} % if too long for running head

%\affil[1]{School of Physics, \protect\\
%          University of Melbourne, \protect\\
%          VIC, Parkville, 3010, Australia}
%\affil[2]{School of Mathematics and Statistics, \protect\\
%          University of Melbourne \protect\\
%          VIC, Parkville, 3010, Australia}
%\affil[3]{Center for Quantum Computation and Communication Technology, \protect\\
%          University of Melbourne \protect\\
%          VIC, Parkville, 3010, Australia}

%\date{2021-01-20}
%\date{Received: date / Accepted: date}
% The correct dates will be entered by the editor

\begin{abstract} \label{sec:abstract}
For universal quantum computation, a major challenge to overcome for practical implementation is the large amount of resources required for fault-tolerant quantum information processing. An important aspect is implementing arbitrary unitary operators built from logical gates within the quantum error correction code. A synthesis algorithm can be used to approximate any unitary gate up to arbitrary precision by assembling sequences of logical gates chosen from a small set of universal gates that are fault-tolerantly performable while encoded in a quantum error-correction code. However, current procedures do not yet support individual assignment of base gate costs and many do not support extended sets of universal base gates. We analysed cost-optimal sequences using an exhaustive search based on Dijkstra’s pathfinding algorithm for the canonical Clifford+$T$ set of base gates and compared them to when additionally including $Z$-rotations from higher orders of the Clifford hierarchy. Two approaches of assigning base gate costs were used. First, costs were reduced to $T$-counts by recursively applying a $Z$-rotation catalyst circuit. Second, costs were assigned as the average numbers of raw (i.e. physical level) magic states required to directly distil and implement the gates fault-tolerantly. We found that the average sequence cost decreases by up to~$54\pm 3\%$ when using the $Z$-rotation catalyst circuit approach and by up to $33\pm 2 \%$ when using the magic state distillation approach. In addition, we investigated observed limitations of certain assignments of base gate costs by developing an analytic model to estimate the proportion of sets of $Z$-rotation gates from higher orders of the Clifford hierarchy that are found within sequences approximating random target gates.
\end{abstract}
\maketitle
\section{Introduction} \label{sec:introduction}
Quantum computing has the potential to solve many real-world problems by using significantly fewer physical resources and computation time than the best known classical algorithms. The quantum algorithms for these problems are implemented using deep quantum circuits. Thus to reliably implement these circuits, qubits within the devices require long coherence times and high precision control. Current systems consist of physical qubits that are too noisy for large scale computation. Error-correction schemes provide the ability to overcome this hurdle by entangling clusters of physical qubits in such a way that they collectively encode the information into more robust logical qubits. In principle, when physical qubits have error-rates below the \textit{error threshold} of the error-correction scheme, logical qubits within the code can be made arbitrarily robust using increasing numbers of qubits. A particular error-correction scheme with relatively high physical error threshold of approximately $1\%$ is the \textit{surface code}, which is implemented over a nearest-neighbour two-dimensional physical layout, making it one of the most realistically implementable schemes~\cite{bravyi1998quantum,dennis2002topological,raussendorf2007topological,wang2011surface}. In this work, we analyse the resource costs for gate synthesis, which is used to fault-tolerantly implement arbitrary unitary gates in error-correction codes.

The surface code, among other high-threshold codes, is limited to a small set of Clifford gates over logical qubits that can be performed with relative ease. A procedure called magic state distillation can be used to perform a wider range of non-Clifford gates fault-tolerantly, such as the $T := R_z(\pi/4)$ gate (up to global phase), which cannot be produced using only Clifford gates~\cite{eastin2009restrictions,zhou2000methodology}. Initially, raw magic states are surgically injected into the code and with the aid of state distillation procedures, a number of raw magic states are consumed to produce a smaller number of more robust magic states. In principle, the procedures can be recursively applied to obtain states with arbitrarily low noise, although requiring large amounts of physical resources. These purified magic states can then be consumed to fault-tolerantly perform corresponding gates using quantum teleportation circuits. Distillation procedures only exist for a subset of gates, in order to implement arbitrary unitary gates, the Solovay-Kitaev (SK) theorem can be used. The SK theorem states that a universal set of~$n$-qubit gates generate a group dense in $SU(2^n)$ (Special Unitary), and the set fills $SU(2^n)$ relatively quickly. Hence single-qubit base gates that form a universal set can be multiplied in sequence to approximate any single-qubit gate to arbitrary precision~\cite{nielsen_chuang_2010, kitaev2002classical}. 

A frequently used set of single-qubit universal base gates for fault-tolerant quantum computation are the Clifford+$T$ gates, where the Clifford gates are relatively cheap to apply while the $T$ gate requires a considerable amount of resources due to the magic state distillation procedure. This set of gates and how they can be used to synthesise arbitrary single-qubit gates is a well studied topic within the quantum compilation literature. Gate synthesis algorithms, besides brute-force~\cite{fowler2011constructing}, began with the Solovay-Kitaev algorithm~\cite{dawson2005solovay, kitaev2002classical}. It initially searches for a base sequence that roughly approximates a target gate and then uses a recursive strategy to append other base sequences in such a way that the new sequence approximates a gate that is closer to the target gate with distance reducing efficiently with the number of iterations. It is compatible with arbitrary single-qubit universal gate sets, provided that they include each gate's adjoint. The SK algorithm has room for optimisation with respect to lengths of resulting gate sequences since the recursive process generates strings of disjoint subsequences which are only individually optimised, rather than optimising over the entire sequence. In~2008, Matsumoto and Amano~\cite{matsumoto2008representation} developed a normal form for sequences of Clifford+$T$ gates that produces unique elements in~$SU(2)$. Shortly after, Bocharov and Svore~\cite{bocharov2012resource} introduced their canonical form which extends the normal form by instead producing unique elements in $PSU(2)$ (Projective Special Unitary) which more concisely describes the space of all physical single-qubit gates by ignoring global phase. This normal form can be used to enumerate length optimal sequences of Clifford+$T$ base gates which produce distinct gates, considerably reducing the size of the sequence configuration space for search algorithms (although still growing exponentially with respect to sequence length). 

More recently, there has been significant progress on developing direct synthesis methods which are not based on search. For target single-qubit unitary gates that can be exactly produced by Clifford+$T$ base gate sequences, a method was developed that optimally and efficiently finds these exact sequences directly~\cite{kliuchnikov2012fast}. This was later used as a subroutine in algorithms for optimal synthesis of arbitrary single-qubit $Z$-rotations~\cite{kliuchnikov2016practical,ross2016optimal}. Direct Clifford+$T$ base gate synthesis methods for $Z$-rotations have since been generalised to Clifford+cyclotomic ($Z$-rotation by $\pi/n$) sets of base gates~\cite{forest2015exact} and sets derived from totally definite quaternion algebras~\cite{kliuchnikov2015framework}. For arbitrary single-qubit rotations (not necessarily $Z$-rotations) there has been a number of other approaches developed, such as a randomised algorithm that uses the distribution of primes~\cite{selinger2012efficient}, asymptotically optimal synthesis using ancilla qubits~\cite{kliuchnikov2013asymptotically}, and probabilistic quantum circuits with fallback~\cite{bocharov2015efficient}.

It is common within the quantum compilation literature for synthesis algorithms to optimise sequences based on minimising the total number of gates that require magic state injection. This measure is well-suited to the Clifford+$T$ set of base gates which are standard for gate synthesis algorithms, since the $T$ gate and its adjoint are the only gates with a significantly higher cost than the Clifford gates. However, procedures exist for performing alternative gates to the $T$ gate that vary in implementation cost. Examples of such gates are found within the Clifford hierarchy, which is an infinite discrete set of gates that are universal and can be performed on certain error-correcting codes fault-tolerantly~\cite{gottesman1999demonstrating}. The resource cost of implementation typically varies between orders of the hierarchy. Thus to accurately cost optimise sequences from such sets of gates, the cost of each individual base gate should be considered. We investigate two different approaches for implementing $Z$-rotation gates from the Clifford hierarchy and calculating their resource costs. The first approach is based on a circuit that uses a catalyst $Z$-rotation state to implement two copies of its corresponding $Z$-rotation gate using a small number of $T$ gates while retaining the initial $Z$-rotation state~\cite{gidney2019efficient, gidney2018halving}. This circuit can enable the average resource costs of implementing $Z$-rotation gates from the Clifford hierarchy to be expressed as $T$-counts. Using this approach, costs could be calculated either by assuming that output gates are applied directly to target qubits or by assuming that all output gates are first applied to $|+\rangle$ states to form intermediate magic states, which can then be consumed to implement the corresponding gates onto target qubits at any time. As an alternative to the $Z$-rotation catalyst circuit approach of gate implementation, the second approach is to use the average number of raw magic states required to directly distil and implement subsets of gates belonging to the Clifford hierarchy in surface codes. The distillation costs have already been calculated by Campbell and O'Gorman~\cite{campbell2016efficient} for various levels of precision, the accumulated costs of distilling and then implementing the gates are found within their supplementary materials. Although other factors relating to physical resources are important to consider such as qubit count, circuit depth, magic state distillation methods, and details of the error-correction implementation, the number of raw magic states can serve as a rough approximation to the cost of implementing fault-tolerant logical gates on surface codes.

We introduce an algorithm, based on Dijkstra's shortest path algorithm, that generates a database of all cost-optimal sequences below a chosen maximum sequence cost where each sequence produces distinct gates in $PSU(2)$. The algorithm supports arbitrary universal sets of single-qubit base gates with individually assigned cost values. The database can then be searched to find a sequence approximating a specified target gate. We use this algorithm to compare the cost of cost-optimal gate synthesis between the canonical Clifford+$T$ base gate set and various sets of base gates consisting of Clifford gates and $Z$-rotations from higher orders of the Clifford hierarchy. Each set of logical base gates is compared by calculating how the average gate sequence cost for approximating random target gates scales with respect to reaching target gate synthesis logical error rates. When including $Z$-rotation base gates from higher orders of the Clifford hierarchy with $T$-counts assigned using the $Z$-rotation catalyst approach, we find that the average cost-optimal sequence $T$-counts can potentially be reduced by over 50\% when output gates are directly applied to target qubits and by over $30\%$ when intermediate magic states are used. When using the alternative approach of assigning costs from direct magic state distillation, we find that by including $Z$-rotation logical base gates from the fourth order of the Clifford hierarchy, the average cost-optimal sequence costs can be reduced by 30\%. These cost reductions indicate that a significant amount of resources could be saved by adapting current synthesis algorithms to include higher orders of the Clifford hierarchy and to optimise sequences with respect to individual gate costs. 

In the cases when costs are assigned using the $Z$-rotation catalyst method via intermediate magic states or when assigned using direct magic state distillation, we observe that there is only a small improvement to the average costs of synthesis when $Z$-rotations of orders higher than four of the Clifford hierarchy are included as base gates. We investigate this behaviour by developing a model to estimate the proportion of $Z$-rotation base gates from specified orders of the Clifford hierarchy within sequences approximating random target gates, without needing to generate the database of sequences. The proportions calculated in this manner closely fit results obtained using the sequence generation algorithm to approximate uniformly distributed random target gates. The parameters of the calculation include the maximum sequence cost and separate logical base gate costs for each order of the Clifford hierarchy, which can be readily be extended to specify costs for individual logical base gates.

\section*{Results}

\subsection*{Base Gates From The Clifford Hierarchy}
\begin{figure*}
	\centering
	\includegraphics[width=1\textwidth]{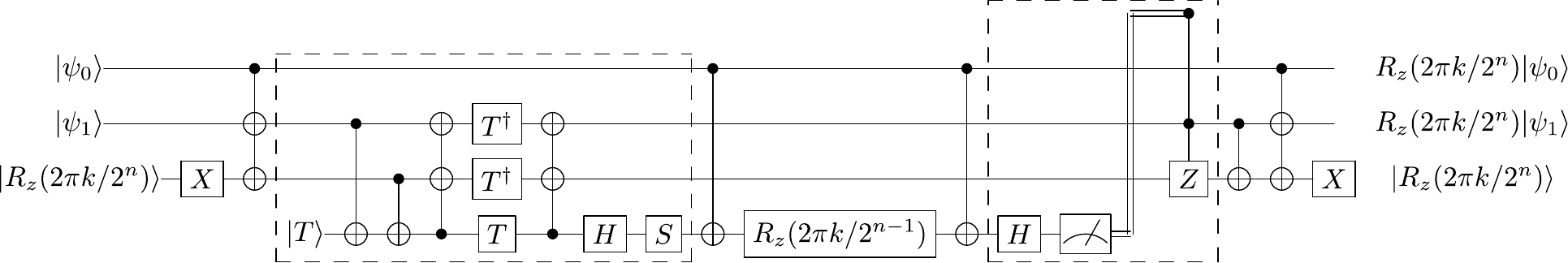}
	\caption{A $Z$-rotation catalyst circuit~\cite{gidney2018halving, gidney2019efficient}. The rotations $R_z(2\pi k / 2^n)$ are elements of $\mathcal{T}_n$ (as shown in Eq.~\ref{eq:z-rotation-base-gate-sets}) where~$k$ is an odd integer and~$n$ is a natural number. The circuit utilises a $|\mathcal{T}_n \rangle$ state, a $|T \rangle$ state, three $\mathcal{T}_{3}$ gates and a $\mathcal{T}_{n-1}$ gate to perform two $\mathcal{T}_{n}$ gates on two separate qubits while retaining the original $|\mathcal{T}_n \rangle$ state. The output $\mathcal{T}_{n}$ gates can either be applied directly to target qubits or $|\psi_0\rangle$ and $|\psi_1\rangle$ states can be first set to $|+\rangle$ states, so that the application of the $\mathcal{T}_n$ gates prepare two $|\mathcal{T}_n\rangle$ states which can then be used to implement $\mathcal{T}_n$ gates at any time and on any target qubit using teleportation circuits. However, this consumes on average an additional half a $\mathcal{T}_{n-1}$ gate for the implementation of each $\mathcal{T}_{n}$ gate. The two sets of grouped gates (outlined by dashed lines) correspond to logical-AND computation and uncomputation circuits, which only requires a total $T$-count of four to implement~\cite{gidney2018halving}. The circuit can be recursively applied until the $R_z(2\pi k/2^{n-1})$ gate position reduces down to a $\mathcal{T}_{3}$ gate which has a cost of 1. All costs are calculated by assuming that all target gates at each recursive level of the circuit are used at some point (i.e. that no output gates are wasted). } \label{fig:z-rotation-catalyst-circuit}
\end{figure*}

The Clifford hierarchy is an infinite discrete set of gates that are universal for the purposes of quantum computation and can be fault-tolerantly performed on certain error-correcting codes. Each order of the hierarchy is defined as
\begin{equation}
C_l := \{U \;|\; UPU^\dag \in C_{l-1}, \;\forall P \in \mathcal{P} \},
\end{equation}
noting that $C_1 = \mathcal{P}$ is the set of Pauli gates, $C_2$ is the set of Clifford gates and $C_3$ includes, among others, the Pauli basis rotations by~$\pi/4$ such as the $T$ gate. Higher order gates typically correspond to finer angle rotations.

In this work, we compare sets of single-qubit universal logical base gates consisting of Clifford gates and $Z$-rotation gates from higher orders of the Clifford hierarchy. Although only higher order $Z$-rotations are included, they can be readily converted to other gates in the same order of the Clifford hierarchy by multiplying gates from lower orders. In particular, by multiplying Clifford gates, other gates of the same order are generated for the same cost. For example~$Z.R_z(\pi/4) = R_z(5\pi/4)$ and $H.R_z(\pi/4).H = R_x(\pi/4)$ up to global phase, where $H$ is the Hadamard gate and~$Z$ is the Pauli-$Z$ gate. These sets of logical base gates are compared with respect to the optimal resource costs resulting from gate synthesis for random target gates. Each set of $Z$-rotation gates from order $3 \leq l \leq 7$ of the Clifford hierarchy, denoted $\mathcal{T}_l$, can be written as
\begin{align}
    \mathcal{T}_3 &:= \left\{ R_z\left( \frac{\pi k}{4} \right) \in C_3 \; \mid \; k \in \{-1, 1\} \right\}, \nonumber\\
    \mathcal{T}_4 &:= \left\{ R_z\left( \frac{\pi k}{8} \right) \in C_4 \; \mid \; k \in \{-3, -1, 1, 3\} \right\}, \nonumber\\
    \mathcal{T}_5 &:= \left\{ R_z\left( \frac{\pi k}{16} \right) \in C_5 \; \mid \; k \in \{-7, -5, \ldots, 5, 7\} \right\}, \nonumber\\
    \mathcal{T}_6 &:= \left\{ R_z\left( \frac{\pi k}{32} \right) \in C_6 \; \mid \; k \in \{-15, -13, \ldots, 13, 15\} \right\}, \; \text{and} \nonumber\\
    \mathcal{T}_7 &:= \left\{ R_z\left( \frac{\pi k}{64} \right) \in C_7 \; \mid \; k \in \{-31, -29, \ldots, 29, 31\} \right\}. \label{eq:z-rotation-base-gate-sets}
\end{align}
The five sets of logical base gates used in our analysis are then constructed as
\begin{align}
	\text{Set}_1 &:= C_1 \cup C_2 \cup \mathcal{T}_3, \nonumber\\
	\text{Set}_2 &:= \text{Set}_1 \cup \mathcal{T}_4, \nonumber\\
	\text{Set}_3 &:= \text{Set}_2 \cup \mathcal{T}_5, \nonumber\\
	\text{Set}_4 &:= \text{Set}_3 \cup \mathcal{T}_6, \text{ and}\nonumber\\
	\text{Set}_5 &:= \text{Set}_4 \cup \mathcal{T}_7. \label{eq:base-gate-sets}
\end{align}

\begin{table}
\small
\centering
\subfloat[Direct application of $\mathcal{T}_l$ method][Direct application of $\mathcal{T}_l$ method\label{tab:direct-application-method}]{
        \vspace{9pt}
        \centering
       {\rowcolors{2}{gray!15}{white}
        \begin{tabular}{c|c}
        \rule{2cm}{0pt}&\rule{2cm}{0pt}\\[-\arraystretch\normalbaselineskip]
        \multicolumn{2}{c}{\text{Average $T$-count per base gate}}\\
        \hline
        $\mathcal{T}_3$ & 1 \\
        $\mathcal{T}_4$ & 2.5 \\
        $\mathcal{T}_5$ & 3.25 \\
        $\mathcal{T}_6$ & 3.625 \\
        $\mathcal{T}_7$ & 3.8125
        \end{tabular}
        }
     }
    \qquad\qquad
    \subfloat[Application of $\mathcal{T}_l$ via $|\mathcal{T}_l\rangle$ method][Application of $\mathcal{T}_l$ via $|\mathcal{T}_l\rangle$ method\label{tab:application-via-magic-method}]{
        \vspace{9pt}
        \centering
        {\rowcolors{2}{gray!15}{white}
        \begin{tabular}{c|c}
        \rule{2cm}{0pt}&\rule{2cm}{0pt}\\[-\arraystretch\normalbaselineskip]
        \multicolumn{2}{c}{\text{Average $T$-count per base gate}}\\
        \hline
        $\mathcal{T}_3$ & 1 \\
        $\mathcal{T}_4$ & 3 \\
        $\mathcal{T}_5$ & 5 \\
        $\mathcal{T}_6$ & 7 \\
        $\mathcal{T}_7$ & 9
        \end{tabular}
        }
     }
\vspace{8pt}
\caption{The average number of $T$ gates required to implement a single qubit $Z$-rotation gate from order~$l$ of the Clifford hierarchy~$\mathcal{T}_l$ using the $Z$-rotation catalyst approach. \textbf{(a)} The average $T$-count required to implement $\mathcal{T}_l$ gates by directly applying them to target qubits. The $T$-counts are calculated using the expression Cost[$\mathcal{T}_l]= 4-3\times 2^{3-l}$ as shown in Equation~\ref{eq:t-count-costs}. \textbf{(b)} The average $T$-count required to implement~$\mathcal{T}_l$ gates by applying them via intermediate~$|\mathcal{T}_l\rangle$ states at every level of recursion (since the $Z$-rotation catalyst circuit is recursively applied). The $T$-counts are calculated using the expression Cost[$\mathcal{T}_l]= 1 + 2\times(l-3)$ as shown in Equation~\ref{eq:t-count-costs-via-magic}.\label{tab:t-count-costs}}
\end{table}

\begin{table*}
\small
\centering
{\rowcolors{2}{gray!15}{white}
\begin{tabular}{c|cccc}
\rule{2cm}{0pt}&\rule{2cm}{0pt}&\rule{2cm}{0pt}&\rule{2cm}{0pt}&\rule{2cm}{0pt}\\[-\arraystretch\normalbaselineskip]
\multicolumn{5}{c}{\text{Average raw magic state count per base gate}}\\
\hline
Base gate error rate $\mu$ & $10^{-5}$ & $10^{-10}$ & $10^{-15}$ & $10^{-20}$ \\
\hline
$\mathcal{T}_3$ & 5.1 & 36.2 & 70.4 & 120.1\\
$\mathcal{T}_4$ & 16.7 & 103.1 & 186.5 & 358.7\\
$\mathcal{T}_5$ & 34.8 & 172.7 & 333.2 & 635.8\\
$\mathcal{T}_6$ & 49.0 & 255.8 & 486.1 & 962.2\\
$\mathcal{T}_7$ & 64.7 & 344.8 & 671.5 & 1351.2
\end{tabular}
}
\vspace{8pt}
\caption{The average raw magic state count required for distillation and implementation of corresponding logical base gates, obtained from the supplementary materials of~\cite{campbell2016efficient}. Each column contains the cost of distilling and implementing a logical~$Z$-rotation gate from order $l$ of the Clifford hierarchy~$\mathcal{T}_l$ to below a gate error rate~$\mu$ calculated using the diamond norm. The raw magic state physical level error is assumed to be 0.1\%.\label{tab-costs} }
\end{table*}

Calculating precise resource costs of implementing each gate fault-tolerantly is an extensive task that would need to consider a variety of factors such as qubit count, circuit depth, magic state distillation methods and details of the error-correction implementation. As an approximation for the cost of these logical gates we investigate two approaches of assigning costs to individual $\mathcal{T}_l$ gates, where gates from $C_1$ and $C_2$ are assumed to be free since they can be implemented in a relatively straightforward way. The first approach can associate the costs with the $T$-count, which is used as the standard metric for measuring the costs of gate sequences within the gate synthesis literature. This can be done by using a $Z$-rotation catalyst circuit shown in Fig.~\ref{fig:z-rotation-catalyst-circuit}, which was introduced in~\cite{gidney2018halving} and presented in more detail in~\cite{gidney2019efficient}. The circuit is similar to a synthillation parity-check circuit described in~\cite{campbell2018magic}. It utilises a $|\mathcal{T}_l \rangle$ state and a small number of $T$ gates to perform two $\mathcal{T}_{l}$ gates on two different qubits while retaining the original $|\mathcal{T}_l \rangle$ state. Costs can be calculated by recursively applying this circuit, assuming that all output gates at each recursive level are resourced (i.e. that no output gates are wasted). We calculate the costs using the $Z$-rotation catalyst approach in two ways. The first assumes that output $\mathcal{T}_{l}$ gates are directly applied to target qubits. The recurrence relation for the $T$-counts using this method can be obtained as
\begin{equation}
    \text{Cost}\left[ \mathcal{T}_l \right] = \frac{4 + \text{Cost}\left[ \mathcal{T}_{l-1} \right]}{2},
\end{equation}
where Cost[$\mathcal{T}_3$] = 1. Solving this results in the average number of $T$ gates required to implement a $\mathcal{T}_l$ gate to be expressed as
\begin{equation}
    \text{Cost}\left[ \mathcal{T}_l \right] = 4 - 3\times 2^{3-l},\label{eq:t-count-costs}
\end{equation}
which is enumerated in Table~\ref{tab:direct-application-method} for $3\leq l\leq 7$. The second method of calculating the $T$-count using the $Z$-rotation catalyst approach applies the $\mathcal{T}_{l}$ gates to $|+\rangle$ states, creating corresponding intermediate $|\mathcal{T}_{l}\rangle$ states, which are then consumed to implement the gates via teleportation circuits. The recurrence relation for these costs can be obtained as
\begin{equation}
    \text{Cost}\left[ \mathcal{T}_l \right] = 2 + \text{Cost}\left[ \mathcal{T}_{l-1} \right],
\end{equation}
where Cost[$\mathcal{T}_3$] = 1, resulting in the expression
\begin{equation}
    \text{Cost}\left[ \mathcal{T}_l \right] = 1 + 2\times (l-3)\label{eq:t-count-costs-via-magic}
\end{equation}
which is enumerated in Table~\ref{tab:application-via-magic-method} for $3\leq l\leq 7$. This second method is more expensive since the teleportation circuit that consumes the $|\mathcal{T}_{l}\rangle$ state to implement the $\mathcal{T}_{l}$ gate requires a $\mathcal{T}_{l-1}$ correction gate to be applied 50\% of the time. However, this method is more flexible in implementation since the outputted $|\mathcal{T}_{l}\rangle$ states can be used at any time to implement $\mathcal{T}_{l}$ gates onto any target qubits, enabling more options when instruction scheduling. A realistic employment of the $Z$-rotation catalyst approach would likely benefit from a combination of both direct application of $\mathcal{T}_{l}$ gates and application via their intermediate $|\mathcal{T}_{l}\rangle$ states. For the second approach of assigning resource costs, we use the average number of raw magic states to implement fault-tolerant $\mathcal{T}_l$ gates from direct magic state distillation procedures. Resource costs have already been calculated for $Y$-rotation gates $R_y(2\pi/2^l)$ from the Clifford hierarchy by searching for optimal combinations of various distillation protocols with respect to target gate synthesis error rates~$\epsilon$~\cite{campbell2016efficient}. For integer multiples $R_y(2\pi k/2^l)$, the distillation protocols can be performed identically, hence they can be assigned the same cost. To follow convention, the $Y$-rotation gates are converted to $Z$-rotation gates with the same cost using the relation $R_z(\theta) = HS^\dag R_y(\theta) SH$, since $H$ and $S:=R_z(\pi/2)$ have zero cost due to being elements of $C_2$. These resource costs vary between orders of the Clifford hierarchy and are shown in Table~\ref{tab-costs}.
\subsection*{Sequence Generation Algorithm} \label{sec:algorithm}
In this section, a sequence generation algorithm, based on Dijkstra's algorithm, is developed that generates a database of all cost-optimal single-qubit gate sequences below some maximum cost using arbitrary sets of universal base gates which have individually assigned cost values. We use this algorithm to help study the average cost of cost-optimal gate synthesis when including $Z$-rotation gates from higher orders of the Clifford hierarchy as base gates. Due to the flexibility of this algorithm, it could be used as a subroutine within other synthesis algorithms. For example, it could be used as the base approximation step within the SK algorithm, enabling the SK algorithm to consider individual base gate costs when synthesising target gates.

\begin{figure*}
	\centering
	\includegraphics[width=0.62\textwidth]{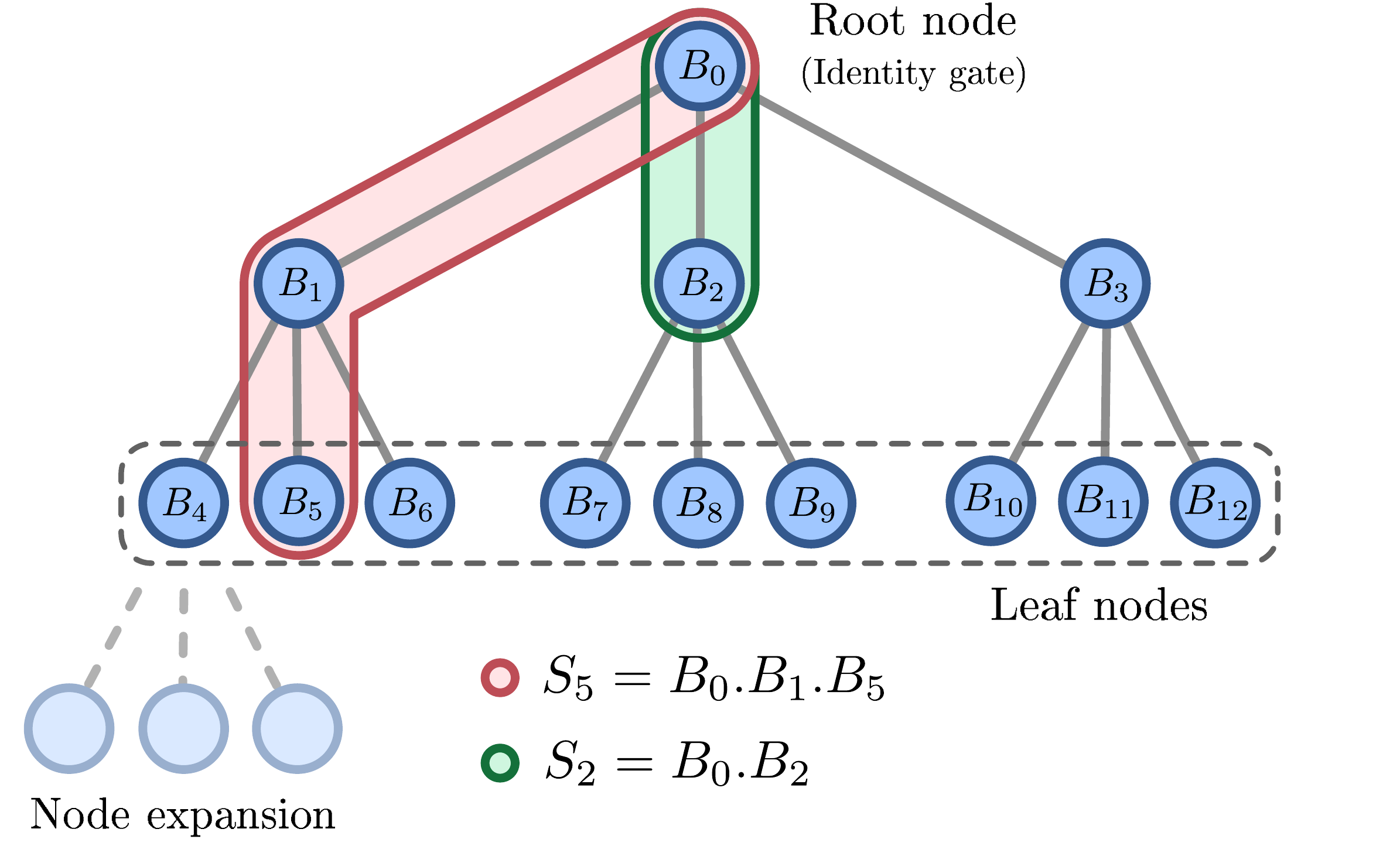}
	\caption{An example of a sequence tree used to relate logical base gates, gate sequences and combined gates for the sequence generation algorithm. A node $n$ corresponds to a single-qubit base gate $B_n$ and the root node corresponds to the identity gate $B_0=I$. A gate sequence corresponding to $n$ is the sequence of logical base gates along the path from $B_0$ to $B_n$. A combined gate $S_n$ is calculated by multiplying all logical base gates within the gate sequence in sequence order. In this example, $B_1$, $B_2$ and $B_3$ are logical base gates where $B_1 = B_4 = B_7 = B_{10}$, $B_2 = B_5 = B_8 = B_{11}$ and $B_3 = B_6 = B_9 = B_{12}$. In the sequence generation algorithm, the leaf node with the lowest sequence cost is expanded by adding a child node as a new leaf node for each gate in the set of logical base gates. All non-leaf nodes of the tree correspond to cost-optimal sequences and they can be thought of as the cost-optimal sequence database generated by the algorithm. Although all leaf nodes are depicted to be at the same depth in the tree, this is not always the case. At any point during the sequence generation algorithm, a path of relatively expensive logical base gates may be much shorter than a path of relatively cheap gates.  }\label{fig:algorithm-graph}
\end{figure*}

The sequence generation algorithm explores the space of sequence configurations using a tree expansion as shown in Figure~\ref{fig:algorithm-graph}, where each node corresponds to a gate and each path from the root node to any other node corresponds to a sequence of gates. Let~$B_n$ be an element of~$PSU(2)$ corresponding to the base gate of node~$n$ in the sequence tree. A \emph{combined gate}~$S_n$ of node $n$ is calculated by multiplying all nodes within the branch from the root down to~$n$, i.e.~$S_n := B_{n_0} \cdot B_{n_1} \ldots B_{n_{k}}$, where~$n_i$ is the~$i^{\text{th}}$ node from the root node such that~$n_0$ is the root and~$n_k$ is node~$n$. The Lie algebra generator of~$S_n$ in the Pauli basis is of the form of a vector~$\alpha_n X + \beta_n Y + \gamma_n Z$ with real coefficients and can be written as~$(\alpha_n, \beta_n, \gamma_n)$. Each vector represents a point in a ball of radius~$\pi/2$ over the Pauli bases~$X$,~$Y$ and~$Z$. Thus each point within the ball is a geometrical location corresponding to a single-qubit gate.

\begin{algorithm}
\small
    \caption{Cost-optimal sequence generation\label{alg:search-algorithm}}
    \begin{algorithmic}[1] % The number tells where the line numbering should start
        \Procedure{GenerateSequences}{baseGates, maxCost}
        	\State {\color{Peach}sequenceDatabase} $\gets$ new KdTree$\langle$Node$\rangle$ \Comment{\footnotesize To store the cost-optimal sequences geometrically\small}
        	\State {\color{NavyBlue}sequenceTree} $\gets$ new Tree$\langle$Node$\rangle$ \Comment{\footnotesize To relate nodes, sequences and combined gates\small}
        	\State {\color{NavyBlue}sequenceTree}.{\color{Mahogany}SetRoot}(Identity gate) \Comment{\footnotesize Set the root node to the identity gate\small}
        	\State {\color{PineGreen}sortedLeafNodes} $\gets$ new MinHeap$\langle$Node$\rangle$ \Comment{\footnotesize To order sequence tree leaf nodes based on sequence cost\small}
        	\State {\color{Orchid}uniqueVectors} $\gets$ new Hashset$\langle$Vector3$\rangle$ \Comment{\footnotesize To test whether sequences have the same combined gates\small}
        	\State Add {\color{NavyBlue}sequenceTree}.root to {\color{PineGreen}sortedLeafNodes}
        	\While{{\color{PineGreen}sortedLeafNodes} not empty}
        		\State $i$ $\gets$ {\color{PineGreen}sortedLeafNodes}.{\color{Mahogany}Pop}() \Comment{\footnotesize Obtains and removes the leaf node with lowest sequence cost\small}
        		\If{{\color{NavyBlue}sequenceTree}.{\color{Mahogany}SequenceCost}($i$) $>$ maxCost}
					\State \textbf{return} {\color{Peach}sequenceDatabase} \Comment{\footnotesize Complete! Ignore $i$ and return cost-optimal sequences\small}
        		\EndIf
        		\State ($\alpha_i, \beta_i, \gamma_i$) $\gets$ {\color{NavyBlue}sequenceTree}.{\color{Mahogany}GetVector}($i$)
        		\If{($\alpha_i, \beta_i, \gamma_i$) not in {\color{Orchid}uniqueVectors}}
					\State Add $i$ to {\color{Peach}sequenceDatabase} \Comment{\footnotesize The node $i$ corresponds to a cost-optimal sequence\small}      			
        			\State Add ($\alpha_i, \beta_i, \gamma_i$) to {\color{Orchid}uniqueVectors}
        			\State childNodes $\gets$ {\color{NavyBlue}sequenceTree}.{\color{Mahogany}GenerateChildren}($i$, baseGates) 
        			\Statex \Comment{\footnotesize Add base gates as child nodes of $i$\small}
        			\ForAll{$j$ in childNodes}
        				\State ($\alpha_j, \beta_j, \gamma_j$) $\gets$ {\color{NavyBlue}sequenceTree}.{\color{Mahogany}GetVector}($j$)
        				\If{($\alpha_j$, $\beta_j$, $\gamma_j$) not in {\color{Orchid}uniqueVectors}}
        					\State Add $j$ to {\color{PineGreen}sortedLeafNodes}
        				\Else
        				    \State Remove $j$ from {\color{NavyBlue}sequenceTree} \Comment{\footnotesize Vector corresponding to childNode $j$ already found\small}
        				\EndIf
        			\EndFor
        		\EndIf
            \EndWhile\label{euclidendwhile}
        \EndProcedure
    \end{algorithmic}
\end{algorithm}
The pseudocode for the algorithm is shown in Algorithm~\ref{alg:search-algorithm}. It works by expanding nodes in a sequence tree (see Figure~\ref{fig:algorithm-graph}). All leaf (end) nodes of the sequence tree are stored in a minimum heap data structure which sorts the leaf nodes based on their corresponding sequence cost in increasing order. This determines the order of nodes to expand. The tree begins as a single identity gate at the root node which is added as the first element to the leaf node heap. At each iteration, the leaf node with the lowest sequence cost, $i$, is taken from the heap, which for the first iteration would be the identity gate node. The vector~$(\alpha_i, \beta_i, \gamma_i)$ is calculated from the combined gate of the corresponding node's sequence. Before expanding a node in the sequence tree, we check whether another node with the same combined gate vector has already been expanded, using a hashset data structure. If the vector exists in the hashset, then the node is removed from the sequence tree and the algorithm proceeds to the next iteration. This repeats until a unique vector is found. When such a vector is found, it is added to the hashset for uniqueness checking in further iterations and the corresponding node in the sequence tree is expanded by generating a child node for each base gate. Each of these child nodes are added to the leaf node heap. To save computation time, adding a child node to the sequence tree and the heap can be limited to when their corresponding vectors are unique. Since vectors of sequences with lower costs are always added to the hashset before those with higher costs, the hashset must only contain vectors corresponding to sequences with the lowest cost among all sequences that produce equivalent combined gates. Thus, whenever a vector is successfully added to the hashset, the corresponding sequence must be cost-optimal. The cost-optimal vector and sequence pair can be stored in a data structure such as a k-d tree which can be used to approximate target gates by geometrically searching for nearest neighbours in the space of vectors. 

There is a notable further optimisation that could be implemented into Algorithm~\ref{alg:search-algorithm}. During the procedure, all non-leaf nodes within the sequence tree correspond to cost-optimal sequences with unique combined gate vectors, that is, each path starting at the root node and ending at any non-leaf node is a shortest path to the sequence's unique combined gate. To see how this could be helpful, first assume that an existing sequence tree needs to grow to a new maximum cost, such that the leaf nodes need to expand multiple times along the same branch. Instead of searching through every combination of base gates as children for a leaf node, the sequence tree itself can be used as a sieve by iterating child nodes from the root that are known to be shortest paths. The tree already contains optimal paths up to a certain depth, so this information could be used to help avoid the tree branches expanding in directions that produce nonoptimal paths to unique combined gates.

In Algorithm~\ref{alg:search-algorithm}, cost-optimal sequences and their corresponding vectors are stored in a k-d tree which uses the Euclidean distance on the vectors to organise the data. Due to the periodic nature of the vectors, there is a small chance of failure in the k-d tree when searching for nearest neighbours to points close to the boundary. With computational overhead, the k-d tree may be modified to help overcome this~\cite{brown2017review}, or a more appropriate data structure such as a vantage point tree \cite{uhlmann1991satisfying,yianilos1993data} may be used instead. In general, further alternative data structures may be used such as the geometric nearest-neighbour access tree \cite{pham2013optimization}.

\subsection*{Synthesis Results}
\begin{figure}
     \centering
     \subfloat[Sequences using the $Z$-rotation catalyst approach with directly applied output gates][Sequences using the $Z$-rotation catalyst approach with directly applied output gates\label{fig:z-rotation-catalyst-method}]{
         \vspace{6pt}
         \centering
         \includegraphics[scale=0.52]{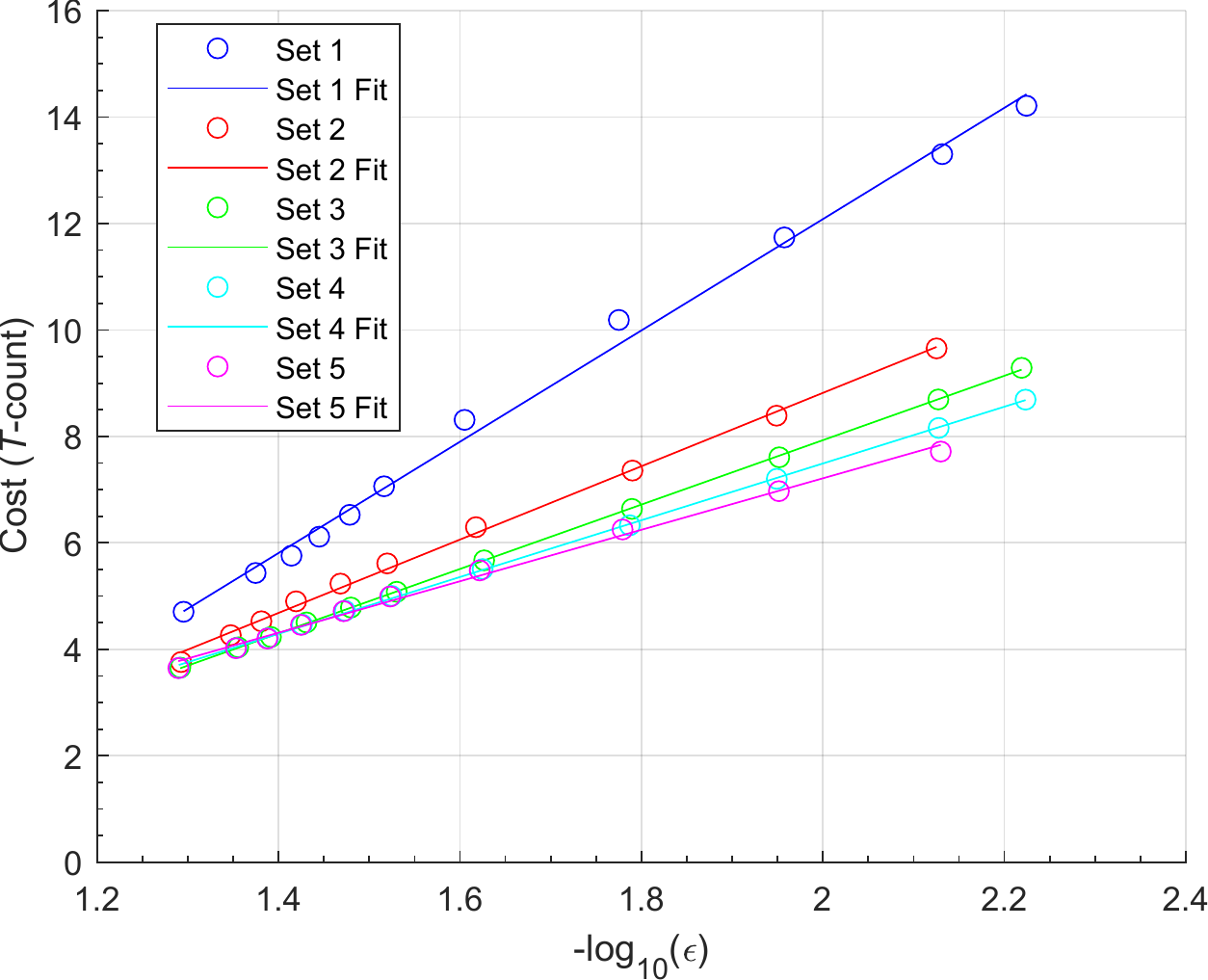}
     }
     \qquad
     \subfloat[Sequences using the $Z$-rotation catalyst approach with output gates applied  via intermediate magic states][Sequences using the $Z$-rotation catalyst approach with output gates applied  via intermediate magic states\label{fig:z-rotation-catalyst-method-for-magic}]{
         \vspace{6pt}
         \centering
         \includegraphics[scale=0.52]{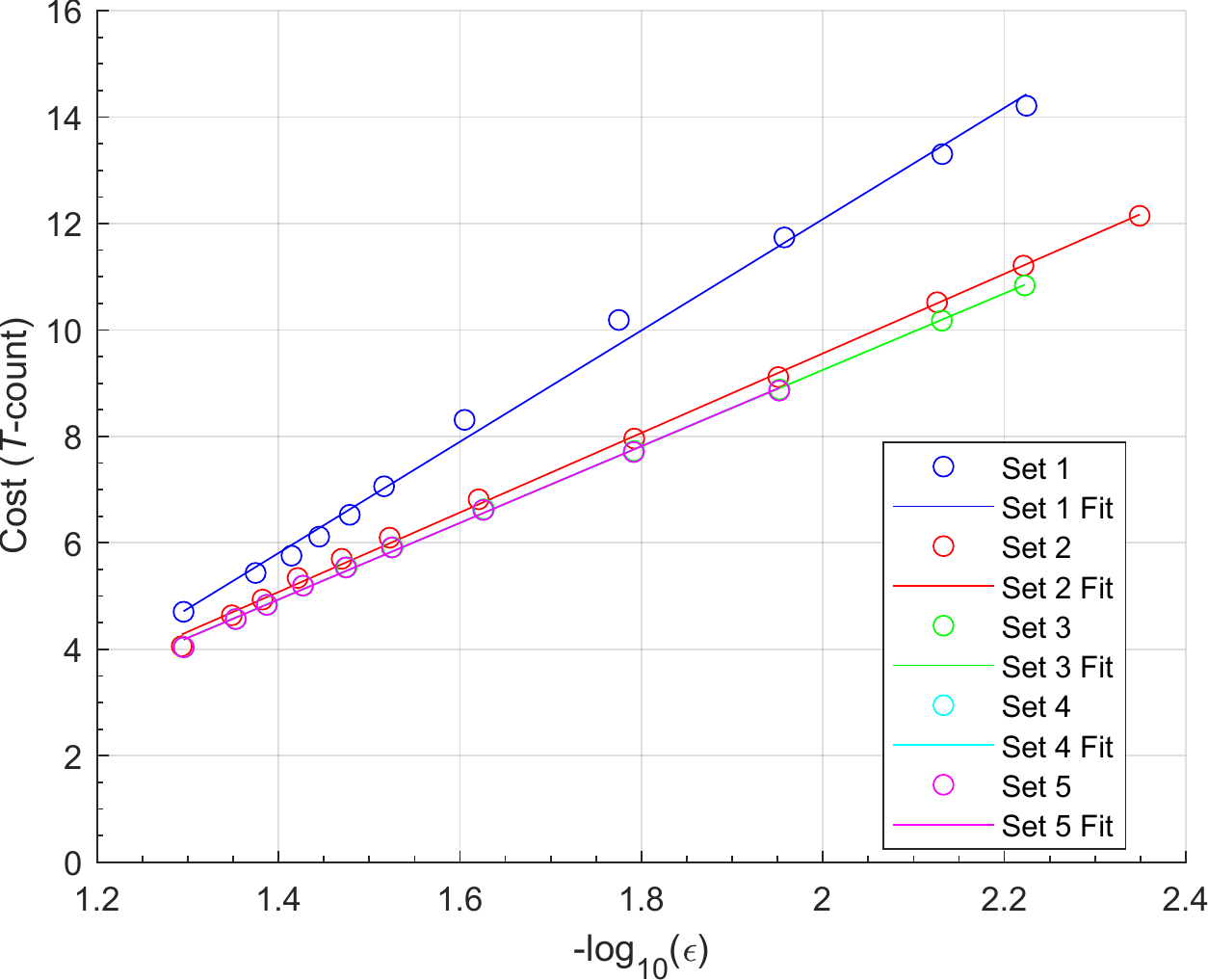}
     }
     \caption{Cost-optimal sequence $T$-counts calculated using the $Z$-rotation catalyst approach plotted against synthesised target gate error rates of~$\epsilon$. Each point is the result of averaging the $T$-count for implementing 5000 random target gate. The synthesis logical errors $\epsilon$ are calculated using the trace distance (shown in Equation~\ref{fig:trace-distance}). The logical base gates for each set of base gates are specified in Equation~\ref{eq:base-gate-sets}. The corresponding linear best fit values for both plots are shown in Table~\ref{tab:t-count-best-fits}. \textbf{(a)} A plot of sequence costs where base gates are assigned costs by assuming that all output gates are directly applied to target qubits. Base gate costs are calculated using Eq.~\ref{eq:t-count-costs} and enumerated in Table~\ref{tab:direct-application-method}. The reductions in scaling factors relative to Set$_1$ are $34\pm 4\%$, $43\pm 2\%$, $49\pm 2\%$, and $54\pm 3\%$ for Set$_2$, Set$_3$, Set$_4$, and Set$_5$ respectively, where uncertainties correspond to~95\% confidence intervals.  These correspond to synthesis cost savings in the limit of small~$\epsilon$. \textbf{(b)} A plot of sequence costs where base gates are assigned costs by assuming that output gates are applied to $|+\rangle$ states to form the corresponding intermediate magic states, gates are then applied by consuming the magic states via teleportation circuits. Base gate costs are calculated using Eq.~\ref{eq:t-count-costs-via-magic} and enumerated in Table~\ref{tab:application-via-magic-method}. The reductions in scaling factors relative to Set$_1$ are $29\pm 3\%$, $31\pm 3\%$, $31\pm 4\%$, and $31\pm 4\%$ for Set$_2$, Set$_3$, Set$_4$, and Set$_5$ respectively, where uncertainties correspond to~95\% confidence intervals.  These correspond to synthesis cost savings in the limit of small~$\epsilon$.}\label{fig:t-count-synthesis-costs}
\end{figure}

\begin{figure}
     \centering
     %\begin{subfigure}[a]{0.49\textwidth}
     %    \centering
     %    \subcaption[short for lof]{Sequences with below $\mu = 10^{-5}$ logical base gate error}\label{fig:cost-5}
     %    \includegraphics[scale=0.55]{synthesis-cost-5.pdf}
     %\end{subfigure}
     %\hfill
     %\begin{subfigure}[a]{0.49\textwidth}
     %    \centering
     %    \subcaption[short for lof]{Sequences with below $\mu = 10^{-10}$ logical base gate %error}\label{fig:cost-10}
     %    \includegraphics[scale=0.55]{synthesis-cost-10.pdf}
     %\end{subfigure}
     %\hfill
     %\begin{subfigure}[a]{0.49\textwidth}
     %    \vspace{6pt}
     %    \centering
     %    \subcaption[short for lof]{Sequences with below $\mu = 10^{-15}$ logical base gate %error}\label{fig:cost-15}
     %    \includegraphics[scale=0.55]{synthesis-cost-15.pdf}
     %\end{subfigure}
     %\hfill
     %\begin{subfigure}[a]{0.49\textwidth}
     %    \vspace{6pt}
     %    \centering
     %    \subcaption[short for lof]{Sequences with below $\mu = 10^{-20}$ logical base gate %error}\label{fig:cost-20}
     %    \includegraphics[scale=0.55]{synthesis-cost-20.pdf}
     %\end{subfigure}
     \subfloat[Sequences with below $\mu = 10^{-5}$ logical base gate error][Sequences with below $\mu = 10^{-5}$ logical base gate error\label{fig:cost-5}]{
         \centering
         \includegraphics[scale=0.50]{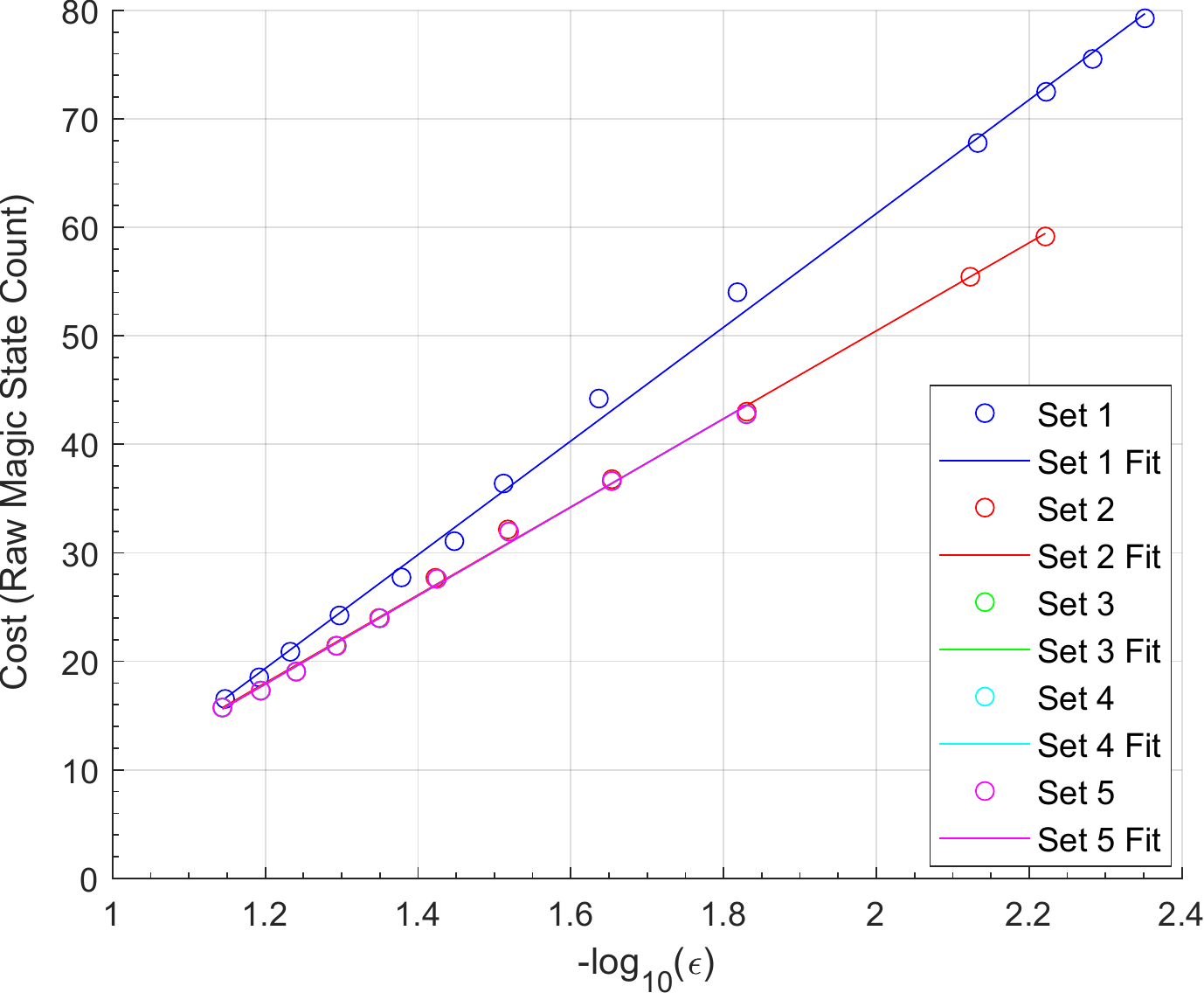}
     }
     \qquad
     \subfloat[Sequences with below $\mu = 10^{-10}$ logical base gate error][Sequences with below $\mu = 10^{-10}$ logical base gate error\label{fig:cost-10}]{
         \centering
         \includegraphics[scale=0.50]{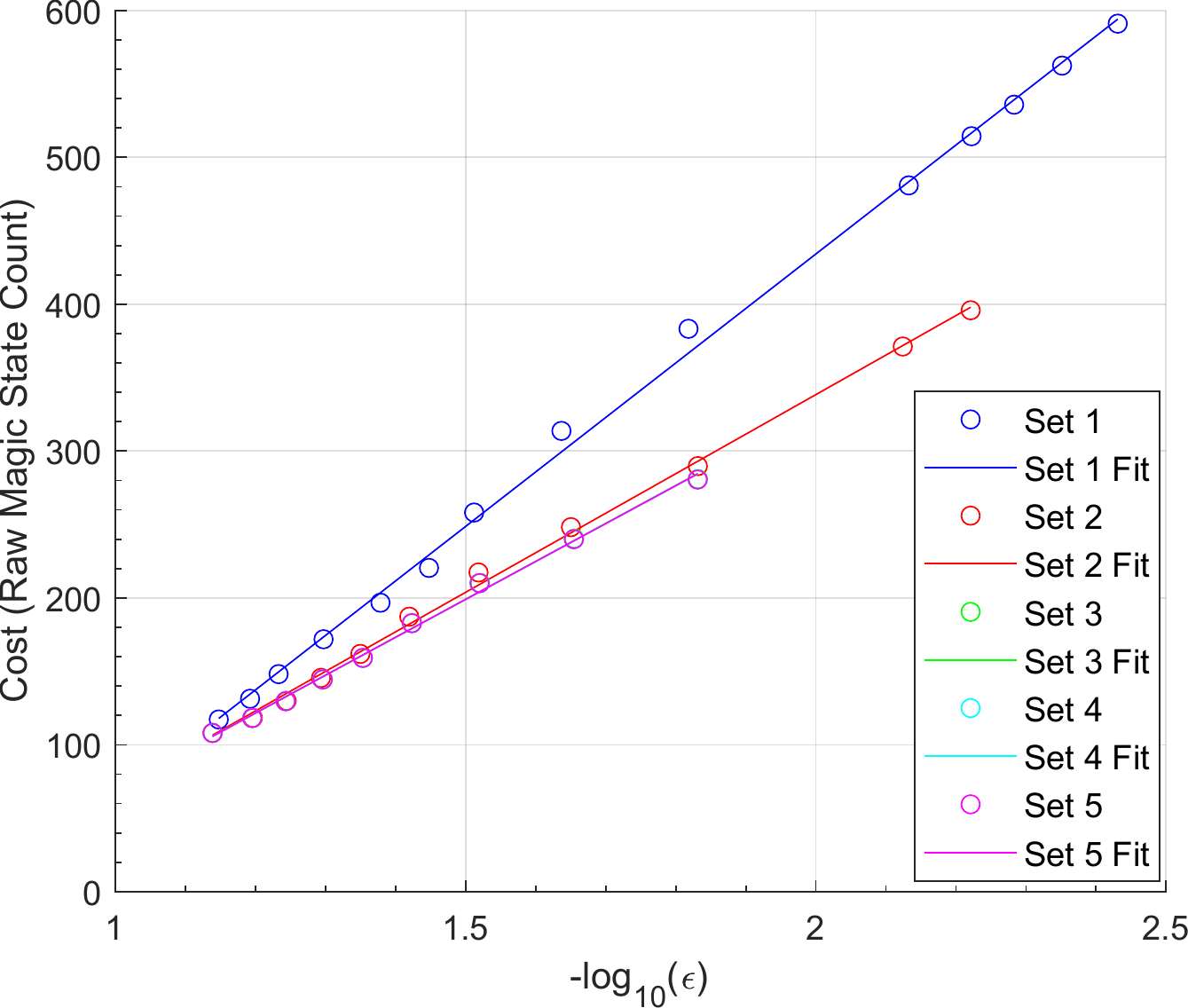}
     }
     \qquad
     \subfloat[Sequences with below $\mu = 10^{-15}$ logical base gate error][Sequences with below $\mu = 10^{-15}$ logical base gate error\label{fig:cost-15}]{
         \vspace{6pt}
         \centering
         \includegraphics[scale=0.50]{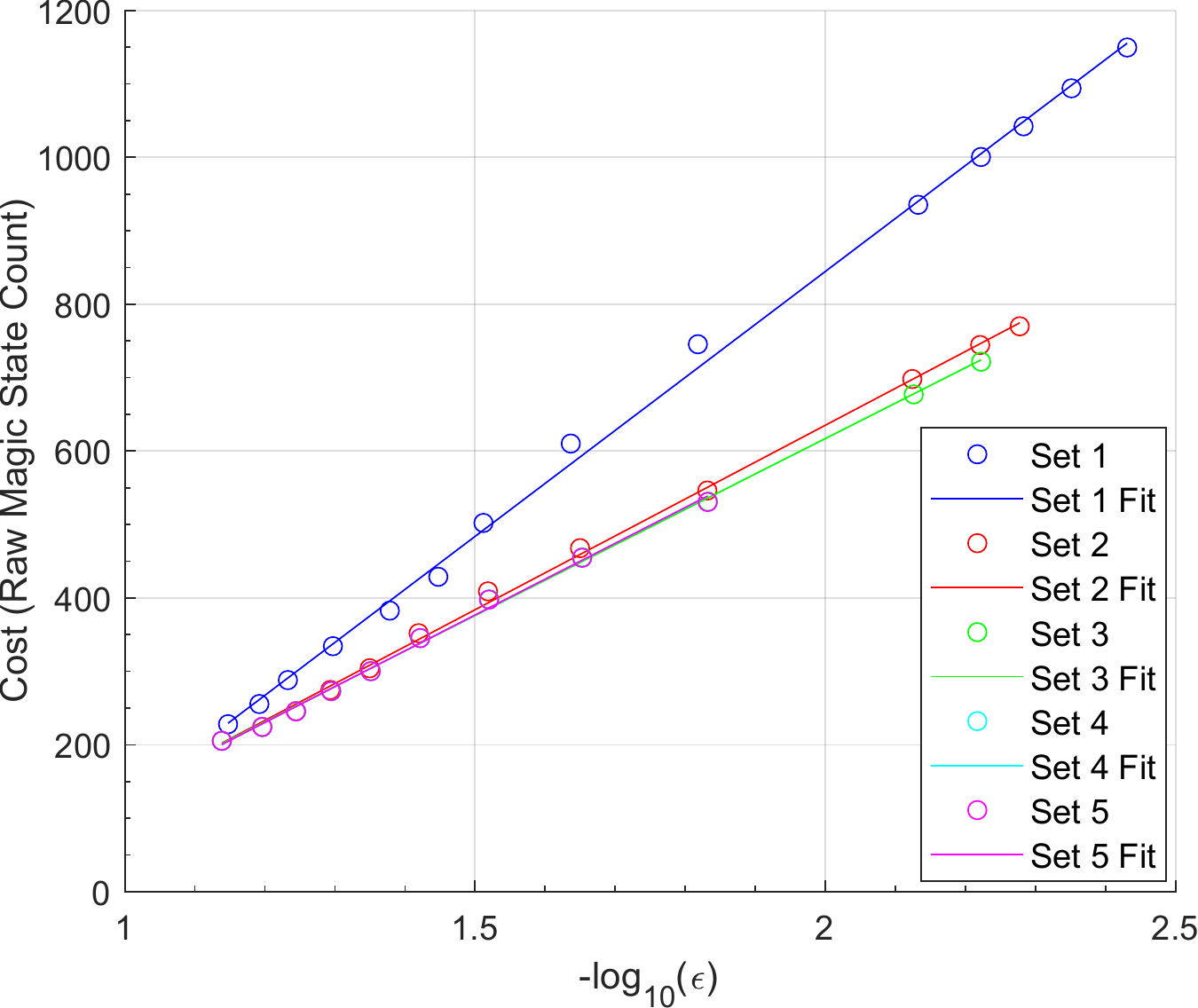}
     }
     \qquad
     \subfloat[Sequences with below $\mu = 10^{-20}$ logical base gate error][Sequences with below $\mu = 10^{-20}$ logical base gate error\label{fig:cost-20}]{
         \vspace{6pt}
         \centering
         \includegraphics[scale=0.50]{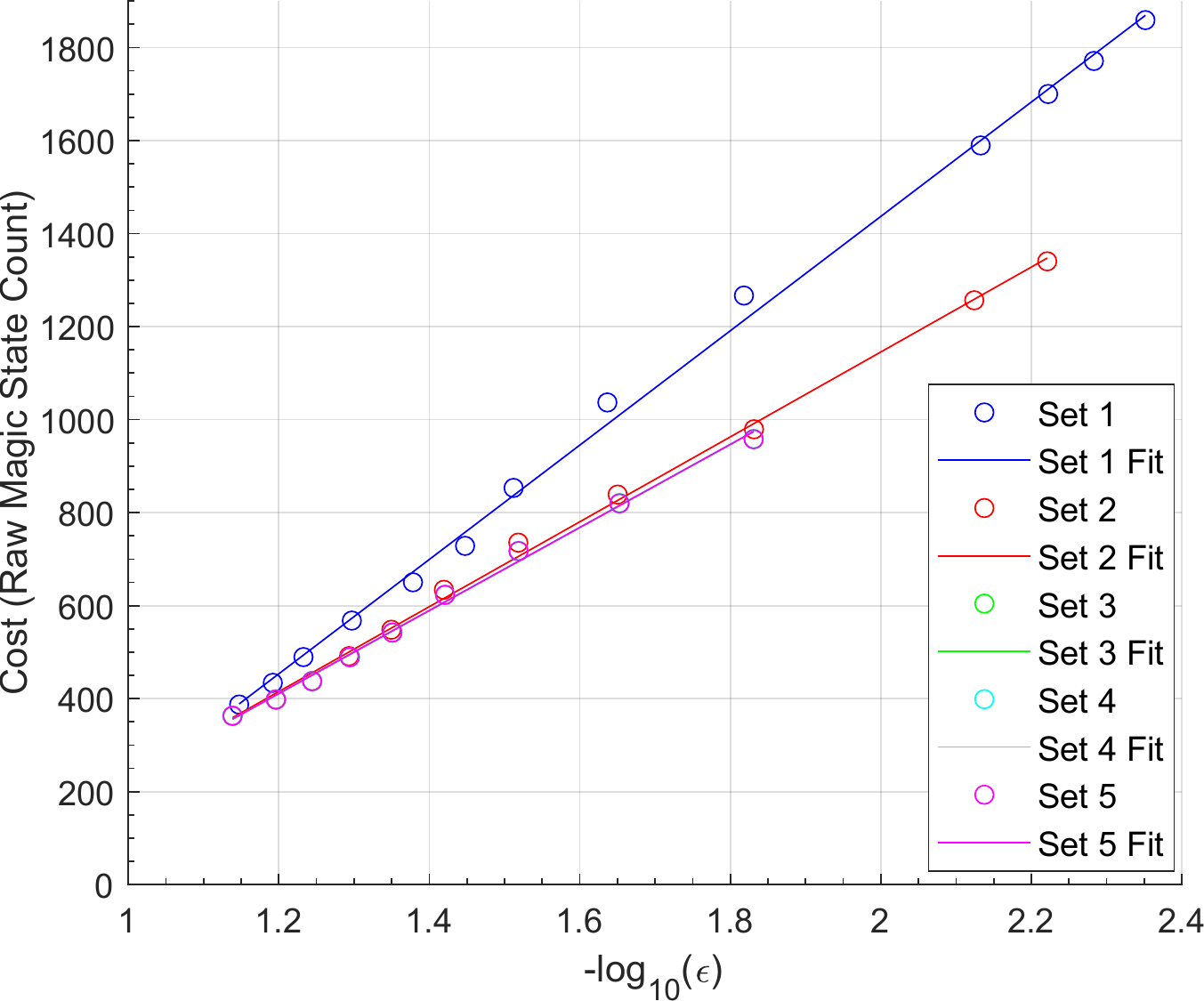}
     }
        \caption{Cost-optimal sequence costs averaged over 5000 random target gates with respect to target gate synthesis logical error rates~$\epsilon$. The logical base gates used are specified in Eq.~\ref{eq:base-gate-sets} with cost values (shown in Table~\ref{tab-costs}) assigned as the average number of raw magic states required to distil and implement them to below a specified logical gate error. The synthesis logical errors $\epsilon$ are calculated using the trace distance (shown in Equation~\ref{fig:trace-distance}). Corresponding linear best fit values are shown in Table~\ref{tab:best-fits}. The pattern of the data about the lines of best fit for each logical base gate set are similar between plots because for each of the logical base gate errors, the ratios of the base gate cost values between orders of the Clifford hierarchy are similar, hence the cost optimal sequences will be comparable. \textbf{(a)} Synthesis using logical base gate costs associated with~$\mu = 10^{-5}$ logical gate error. \textbf{(b)} Synthesis using logical base gate costs associated with~$\mu = 10^{-10}$ logical gate error. \textbf{(c)} Synthesis using logical base gate costs associated with~$\mu = 10^{-15}$ logical gate error. \textbf{(d)} Synthesis using logical base gate costs associated with~$\mu = 10^{-20}$ logical gate error.}\label{fig:synthesis-costs}
\end{figure}
Algorithm~\ref{alg:search-algorithm} was computed using the sets of logical base gates described in Eq.~\ref{eq:base-gate-sets} with the assignment of costs obtained from the two approaches of implementing base gates, where values are shown in Tables~\ref{tab:t-count-costs} and~\ref{tab-costs}. A database was generated that is in the form of a k-d tree of cost-optimal sequences up to some chosen maximum sequence cost. The sequences were organised in the k-d tree with respect to the vectors corresponding to their combined gates. For a given target gate~$G$, gate synthesis was performed by searching for the lowest cost sequence among all nearest neighbours of~$G$ up to a chosen synthesis error (distance), $\epsilon$, between their combined gates and~$G$. The errors were computed using the trace distance defined as 
\begin{equation}
    \mathrm{dist}(S,G) = \sqrt{(2 - |\mathrm{tr}(S^\dag G)|)/2}, \label{fig:trace-distance}
\end{equation}
where $S$ is a combined gate and $G$ is the target gate. If such a sequence did not exist, then the database was further generated to a higher cost and the process was repeated until a sequence was found. Incrementally generating the cost-optimal sequence database in this manner helps avoid over generation. 

For each set of base gates with individual costs calculated for each approach of implementing them, gate synthesis was performed on 5000 random target gates sampled from a uniform distribution for a variety of synthesis error rates~$\epsilon$ (calculated using Eq.~\ref{fig:trace-distance} with respect to the sequences' combined gates). Cost-optimal sequence $T$-counts calculated using the $Z$-rotation catalyst circuit approach for the two methods of assigning base gate costs are plotted against synthesised target gate error rates for each set of base gates in Figure~\ref{fig:t-count-synthesis-costs}. The corresponding linear best fit values for each set of logical base gates and corresponding cost values are shown in Table~\ref{tab:t-count-best-fits}. We can compare the scaling factors of the fits between different sets of logical base gates to estimate changes in average sequence costs as the synthesis error $\epsilon$ approaches zero. 
For the $Z$-rotation catalyst circuit method that assumes all output gates are directly applied to target qubits (as opposed to using intermediate magic states), we find cost savings relative to Set$_1$ of $34 \pm 3\%$, $42 \pm 2\%$, $49 \pm 2\%$, and $54 \pm 3\%$ for Set$_2$, Set$_3$, Set$_4$, and Set$_5$ respectively, where uncertainties correspond to 95\% confidence intervals. Data for a Set$_6$ that includes $\mathcal{T}_8$ gates was also calculated, however no noticeable improvement was found with sequence cost values being almost identical to Set$_5$ resulting in a cost saving of $52 \pm 3\%$ relative to Set$_1$. 
For the $Z$-rotation catalyst circuit method that assumes all output gates are applied to $|+\rangle$ states forming intermediate magic states before consuming them to perform the corresponding $Z$-rotation gate, we find cost savings relative to Set$_1$ of $29\pm 3\%$, $31\pm 3\%$, $31\pm 4\%$, and $31\pm 4\%$ for Set$_2$, Set$_3$, Set$_4$, and Set$_5$ respectively. These results show that if gate synthesis includes higher order Clifford hierarchy $Z$-rotation gates as base gates implemented using the $Z$-rotation catalyst approach, then a $T$-count saving of over $50\%$ could potentially be achieved. Cost-optimal sequence raw magic state counts calculated using direct base gate distillation and implementation procedures are plotted against synthesised target error rates for each combination of base gates and cost values in Figure~\ref{fig:synthesis-costs}. Each of the four plots correspond to different resource costs of distilling and implementing the logical base gates with corresponding logical errors $\mu = 10^{-5}$, $10^{-10}$, $10^{-15}$ and $10^{-20}$ calculated using the diamond norm. The corresponding linear best fit values for each set of logical base gates are shown in Table~\ref{tab:best-fits} and corresponding cost values are shown in Table~\ref{tab-costs} (physical error rate assumed to be 0.1\% in all calculations). The pattern of the data about their lines of best fit for each base gate set are similar between plots. This is because for each of the logical base gate errors, the ratios of the logical base gate cost values between orders of the Clifford hierarchy are similar, hence the cost optimal sequences will be comparable. For logical base gate errors $\mu = 10^{-5}$, $10^{-10}$, $10^{-15}$ and $10^{-20}$, we find that Set$_2$ provides~$23 \pm 3 \%$, $27 \pm 2 \%$, $30 \pm 2 \%$ and $26 \pm 3 \%$ reductions in scaling factor respectively compared to Set$_1$. For $\mu = 10^{-10}$ and $10^{-15}$, we find that Set$_3$ provides $30\pm 3\%$ and $33\pm 2\%$ reductions in scaling factor respectively compared to Set$_1$, which are both approximately a further $3\%$ savings compared to Set$_2$. No further improvements are noticeable in our data for these assignments of cost values. These results show that for any error-correction scheme with distillation costs assigned according to Table~\ref{tab-costs}, using Set$_2$ (which includes~$\mathcal{T}_4$ as logical base gates) instead of the standard Set$_1$, reduces the average resource cost scaling factor with respect to the synthesis negative log-error, $\log (\epsilon^{-1})$, by up to~$30\%$. Additionally Set$_3$ can provide up to a further~$3\%$ reduction when compared to~Set$_2$. Each method of assigning individual base gate costs that were used in this work indicated that the resource requirements of synthesis algorithms may be considerably improved by including higher orders of the Clifford hierarchy as logical base gates and by optimising with respect to the individual costs of implementing them.

\begin{table}
\small
     \centering
    \subfloat[Linear fits using $Z$-rotation catalyst method][Linear fits using $Z$-rotation catalyst method\label{tab:t-count-best-fits-direct}]{
        \vspace{9pt}
        \centering
       {\rowcolors{2}{white}{gray!15}
            \begin{tabular}{c|cc}
            Base Gates & Scaling Factor & Constant \\
            \hline
            $\text{Set}_1$ & $10.46 \pm 0.43$ & $-8.83 \pm 0.73$ \\
            $\text{Set}_2$ & $6.89 \pm 0.22$ & $-4.96 \pm 0.36$ \\
            $\text{Set}_3$ & $6.05 \pm 0.03$ & $-4.17 \pm 0.06$ \\
            $\text{Set}_4$ & $5.33 \pm 0.06$ & $-3.18 \pm 0.11$ \\
            $\text{Set}_5$ & $4.84 \pm 0.21$ & $-2.46 \pm 0.34$
            \end{tabular}
            }
     }
    \qquad
    \subfloat[Linear fits using $Z$-rotation catalyst method via magic states][Linear fits using $Z$-rotation catalyst method via magic states\label{tab:t-count-best-fits-via-magic}]{
        \vspace{9pt}
        \centering
        {\rowcolors{2}{white}{gray!15}
            \begin{tabular}{c|cc}
            Base Gates & Scaling Factor & Constant \\
            \hline
            $\text{Set}_1$ & $10.46 \pm 0.43$ & $-8.83 \pm 0.73$ \\
            $\text{Set}_2$ & $7.47 \pm 0.15$ & $-5.39 \pm 0.26$ \\
            $\text{Set}_3$ & $7.19 \pm 0.12$ & $-5.13 \pm 0.21$ \\
            $\text{Set}_4$ & $7.21 \pm 0.25$ & $-5.16 \pm 0.38$ \\
            $\text{Set}_5$ & $7.21 \pm 0.25$ & $-5.15 \pm 0.39$
            \end{tabular}
            }
     }
    \caption{Linear best fits with a confidence level of $95\%$ for cost-optimal sequence costs averaged over random target logical gates with respect to the negative log-error, $\log (\epsilon^{-1})$, for target gate synthesis calculated using the trace distance (shown in Equation~\ref{fig:trace-distance}). The sequences are constructed using logical base gates with cost values assigned according to Table~\ref{tab:t-count-costs}. The corresponding plots are shown in Figure~\ref{fig:t-count-synthesis-costs}.\label{tab:t-count-best-fits}}
\end{table}

\begin{table}
\small
     \centering
     \subfloat[Linear fits for Figure~\ref{fig:cost-5} for below $\mu = 10^{-5}$ logical base gate error][Linear fits for Figure~\ref{fig:cost-5} for below $\mu = 10^{-5}$ logical base gate error\label{tab:best-fits-5}]{
            \centering
            {\rowcolors{2}{white}{gray!15}
            \begin{tabular}{c|cc}
            Base Gates & Scaling Factor & Constant \\
            \hline
            $\text{Set}_1$ & $52.4 \pm 1.3$ & $-43.5 \pm 2.2$ \\
            $\text{Set}_2$ & $40.6 \pm 0.9$ & $-30.7 \pm 1.5$ \\
            $\text{Set}_3$ & $40.8 \pm 1.8$ & $-31.0 \pm 2.6$ \\
            $\text{Set}_4$ & $40.8 \pm 1.8$ & $-31.0 \pm 2.6$ \\
            $\text{Set}_5$ & $40.8 \pm 1.8$ & $-31.0 \pm 2.6$
            \end{tabular}
            }
     }
     \qquad
     \subfloat[Linear fits for Figure~\ref{fig:cost-10} for below $\mu = 10^{-10}$ logical base gate error][Linear fits for Figure~\ref{fig:cost-10} for below $\mu = 10^{-10}$ logical base gate error\label{tab:best-fits-10}]{
        \vspace{9pt}
        \centering
        {\rowcolors{2}{white}{gray!15}
        \begin{tabular}{c|cc}
            Base Gates & Scaling Factor & Constant \\
            \hline
            $\text{Set}_1$ & $371 \pm 8$ & $-308 \pm 14$ \\
            $\text{Set}_2$ & $269 \pm 7$ & $-200 \pm 11$ \\
            $\text{Set}_3$ & $258 \pm 11$ & $-189 \pm 15$ \\
            $\text{Set}_4$ & $258 \pm 11$ & $-188 \pm 15$ \\
            $\text{Set}_5$ & $258 \pm 11$ & $-188 \pm 15$ 
        \end{tabular}
        }
     }
     \\
     \subfloat[Linear fits for Figure~\ref{fig:cost-15} for below $\mu = 10^{-15}$ logical base gate error][Linear fits for Figure~\ref{fig:cost-15} for below $\mu = 10^{-15}$ logical base gate error\label{tab:best-fits-15}]
     {
        \vspace{9pt}
        \centering
        {\rowcolors{2}{white}{gray!15}
        \begin{tabular}{c|cc}
            Base Gates & Scaling Factor & Constant \\
            \hline
            $\text{Set}_1$ & $722 \pm 15$ & $-599 \pm 27$ \\
            $\text{Set}_2$ & $503 \pm 11$ & $-370 \pm 17$ \\
            $\text{Set}_3$ & $482 \pm 10$ & $-347 \pm 16$ \\
            $\text{Set}_4$ & $488 \pm 21$ & $-355 \pm 30$ \\
            $\text{Set}_5$ & $488 \pm 21$ & $-355 \pm 30$
        \end{tabular}
        }
    }
     \qquad
     \subfloat[Linear fits for Figure~\ref{fig:cost-20} for below $\mu = 10^{-20}$ logical base gate error][Linear fits for Figure~\ref{fig:cost-20} for below $\mu = 10^{-20}$ logical base gate error\label{tab:best-fits-20}]
     {
        \vspace{9pt}
        \centering
        {\rowcolors{2}{white}{gray!15}
        \begin{tabular}{c|cc}
            Base Gates & Scaling Factor & Constant \\
            \hline
            $\text{Set}_1$ & $1230 \pm 30$ & $-1020 \pm 50$ \\
            $\text{Set}_2$ & $913 \pm 24$ & $-680 \pm 39$ \\
            $\text{Set}_3$ & $893 \pm 41$ & $-661 \pm 59$ \\
            $\text{Set}_4$ & $893 \pm 41$ & $-661 \pm 59$ \\
            $\text{Set}_5$ & $893 \pm 41$ & $-661 \pm 59$
        \end{tabular}
        }
     }
     %\vspace{8pt}
    \caption{Linear best fits with a confidence level of $95\%$ for cost-optimal sequence costs averaged over random target logical gates with respect to the negative log-error, $\log (\epsilon^{-1})$, for target gate synthesis calculated using the trace distance (shown in Equation~\ref{fig:trace-distance}). The sequences are constructed using logical base gates with cost values assigned according to Table~\ref{tab-costs}. The corresponding plots are shown in Figure~\ref{fig:synthesis-costs}.\label{tab:best-fits}}
\end{table}

\subsection*{Modelling Gate Proportions}\label{sec:modelling-gate-proportions}
\begin{figure}
	\centering
	\subfloat[Proportion of $\mathcal{T}_4$ gates among $\mathcal{T}_3 \cup \mathcal{T}_4$ gates][Proportion of $\mathcal{T}_4$ gates among $\mathcal{T}_3 \cup \mathcal{T}_4$ gates\label{fig:proportions-t4}]
	{
         \centering
         \includegraphics[scale=0.65]{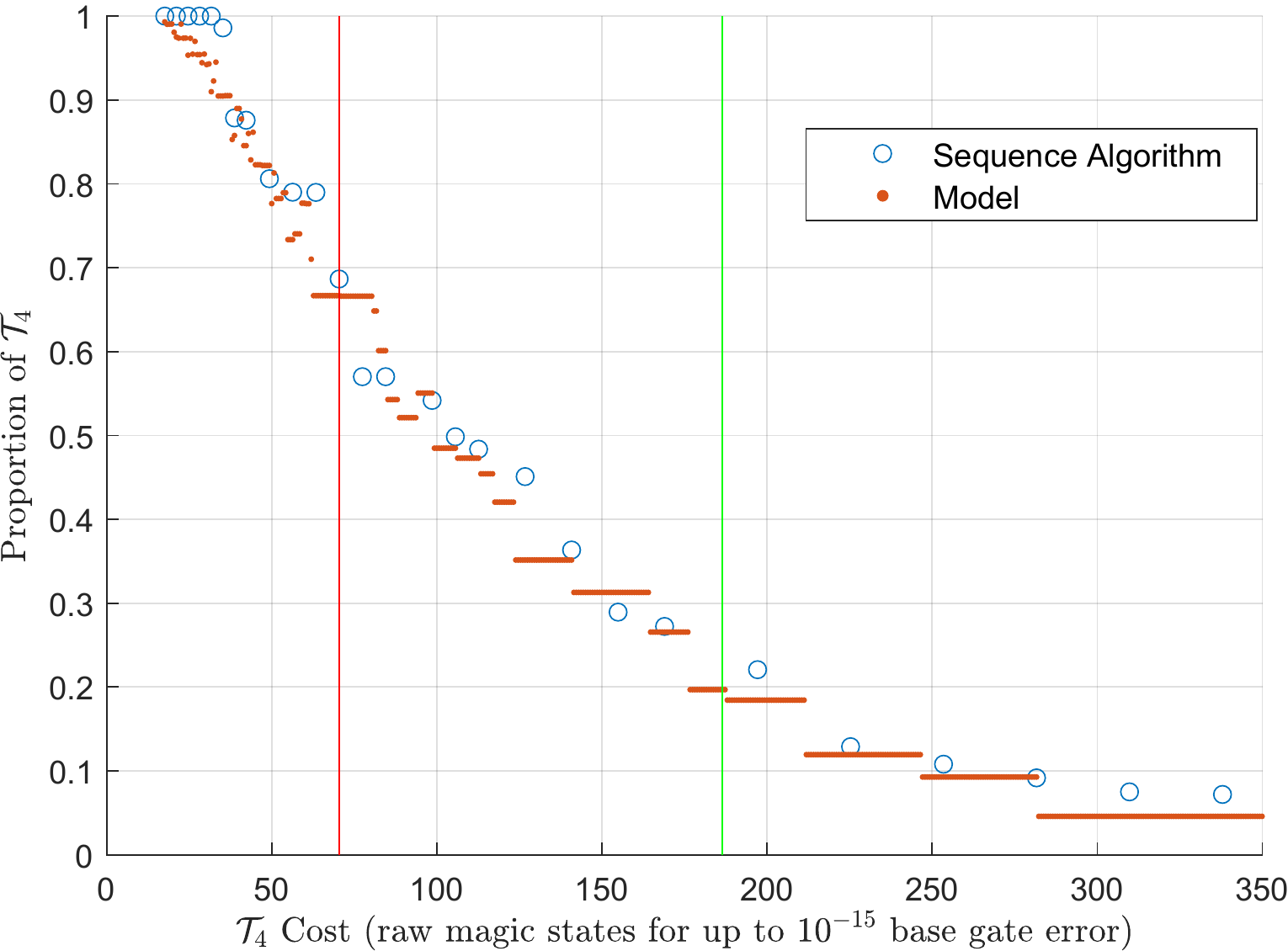}
     }
     \hfill
     \subfloat[Proportion of $\mathcal{T}_5$ gates among $\mathcal{T}_3 \cup \mathcal{T}_4 \cup \mathcal{T}_5$ gates][Proportion of $\mathcal{T}_5$ gates among $\mathcal{T}_3 \cup \mathcal{T}_4 \cup \mathcal{T}_5$ gates\label{fig:proportions-t5}]
     {
         \centering
         \includegraphics[scale=0.65]{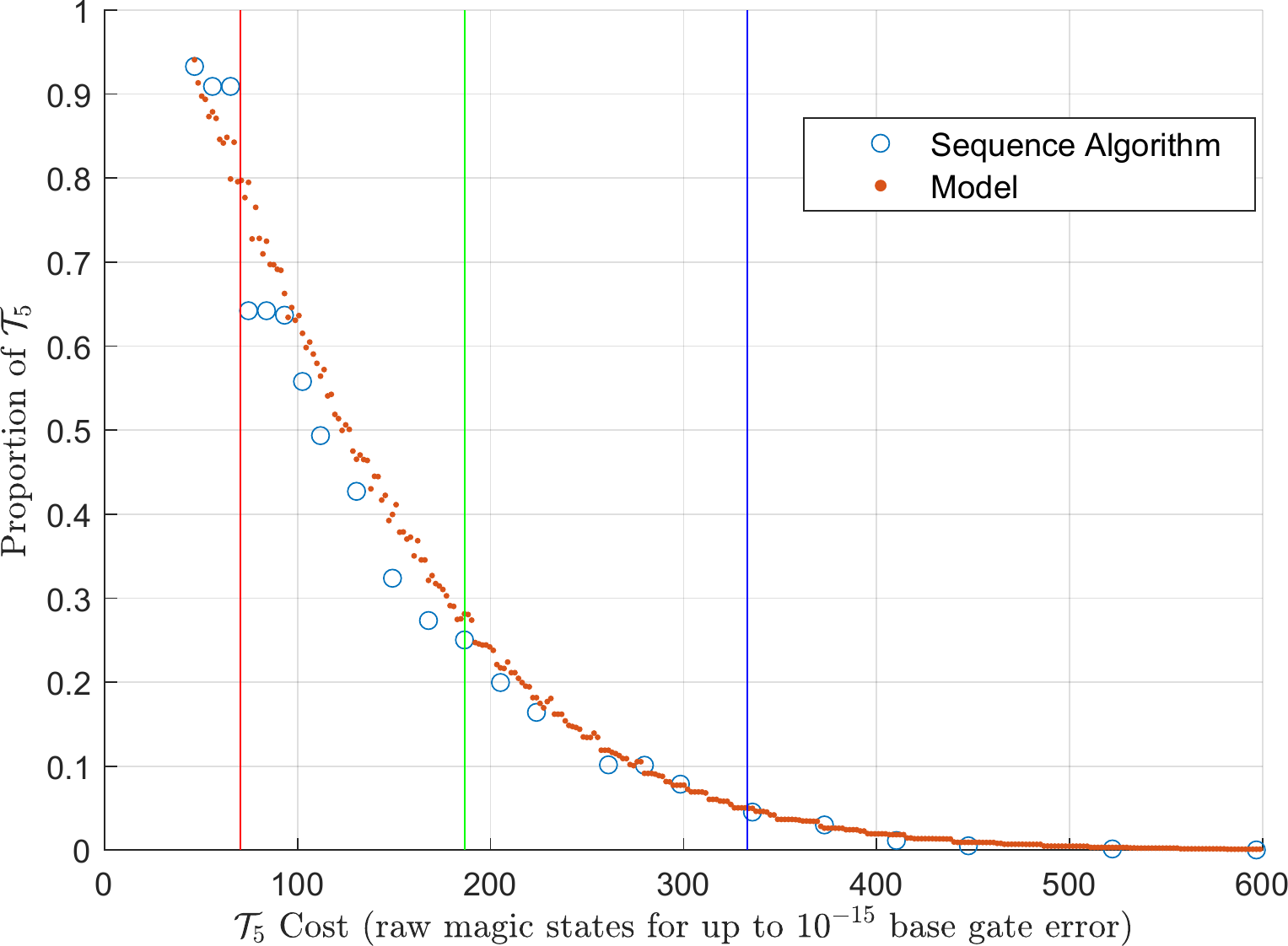}
    }
	\caption{This figure shows the summed proportions of logical base gates from sequences resulting from the sequence generation algorithm and the proportions calculated using our model.
    The sequence generation algorithm outputs cost-optimal sequences approximating random target gates to within~$\epsilon=0.03$ synthesis logical gate error under the trace distance (see Eq.~\ref{fig:trace-distance}), while the model outputs the proportion of a set of logical base gates within the space of all cost-optimal sequences below a maximum cost that produce distinct combined gates. Clifford gates are ignored in the calculations since they are assumed to have zero cost. Both plots show that the model data closely fit the corresponding results from the sequence generation algorithm. The data show that increasing the logical base gate distillation and implementation cost of a particular set~$\mathcal{T}_n$ drastically lowers the proportion of them found within the generated cost-optimal sequences. Thus the set~$\mathcal{T}_n$ with increased costs becomes less effective at reducing the average cost-optimal sequence costs, since they are found less frequently within the sequences. Logical base gate costs are assigned according to Table~\ref{tab-costs} with a logical base gate error of~$\mu = 10^{-15}$ calculated using the diamond norm. The red, green and blue vertical lines (ordered left to right) indicate the logical base gate distillation and implementation costs for~$\mathcal{T}_3$,~$\mathcal{T}_4$ and~$\mathcal{T}_5$ respectively. \textbf{(a)}~The summed proportions of~$\mathcal{T}_4$ logical base gates among~$\mathcal{T}_3 \cup \mathcal{T}_4$ gates for cost-optimal sequences consisting of Set$_2$ logical base gates. Logical base gates from~$\mathcal{T}_3$ are fixed while the cost for~$\mathcal{T}_4$ gates vary. \textbf{(b)} The summed proportions of~$\mathcal{T}_5$ logical base gates among~$\mathcal{T}_3 \cup \mathcal{T}_4 \cup \mathcal{T}_5$ gates for cost-optimal sequences consisting of Set$_3$ logical base gates. Logical base gates from~$\mathcal{T}_3 \cup \mathcal{T}_4$ are fixed while the cost for~$\mathcal{T}_5$ gates vary.\label{fig:proportions}}
\end{figure}

For the raw magic state approach of implementing base gates and the $Z$-rotation catalyst circuit method that uses intermediate magic states, the logical base gate sets Set$_3$, Set$_4$ and Set$_5$ (see Eq.~\ref{eq:base-gate-sets}) were shown to provide only marginal resource savings for gate synthesis when compared with Set$_2$ (see Figs.~\ref{fig:z-rotation-catalyst-method-for-magic} and~\ref{fig:synthesis-costs}), even though the sets contain many more logical base gates. To investigate this behaviour we develop a model in Appendix~\ref{sec:gate-proportions} for determining the proportion of sets of gates among all $\mathcal{T}_l$ gates where~$l \geq 3$ within cost-optimal sequences approximating random target gates with specified gate costs. The proportions can provide insight into how the average sequence cost changes with respect to which $\mathcal{T}_l$ base gates are included as logical base gates and what cost values are assigned. For logical base gates with non-zero proportion within sequences approximating target gates, we expect that by increasing their cost, their recalculated proportion will decrease and the average cost of these sequences will increase. Furthermore, for sets of logical base gates with relatively small proportions, the average sequence cost would only slightly increase if the set were to be excluded compared to sets of base gates with larger proportions.

The model estimates the average proportion,~$p_n$, of $\mathcal{T}_n$ logical base gates among all $\mathcal{T}_l$ gates where $l\geq 3$ from within cost-optimal sequences approximating random target gates to within sufficiently small synthesis errors $\epsilon$. The construction is based on a unique canonical form~\cite{forest2015exact} for sequences of logical base gates and is defined as
\begin{equation}
    c . t_1 . H . t_2 . H \ldots t_{N} . c^\prime,
\end{equation}
where $c$ and $c^\prime$ are Clifford gates, $H$ is the Hadamard gate, $t_m$ is the $m^{\text{th}}$ positioned $Z$-rotation gate from order three and above of the Clifford hierarchy, and~$M$ is the total number of $t_m$ gates in the sequence. This canonical form has the property that arbitrary gate sequences with distinct combined gates, where the sequences can consist of logical base gates from the Clifford gates and $Z$-rotations from orders three and above of the Clifford hierarchy, can be reduced to distinct sequences of this form. The gate proportion for~$\mathcal{T}_n$, denoted $p_n$, can be calculated by averaging the~$\mathcal{T}_n$ logical gate count over all possible sequences in this canonical form that are below a chosen maximum cost $C$ (as detailed in Appendix~\ref{sec:gate-proportions}). That is, 
\begin{align}
    p_n &=\frac{\sum \limits_{k_3=0}^{\lfloor C/c_3 \rfloor} \sum \limits_{k_4=0}^{\lfloor (C - c_3k_3)/c_4 \rfloor} \ldots \sum \limits_{k_L=0}^{\lfloor(C - \sum \limits_{j=3}^{L-1}c_j k_j)/c_L \rfloor}k_n\left(\sum\limits_{i=3}^{L}k_i\right)!\prod \limits_{l=3}^{L} \frac{{|\mathcal{T}_l|}^{k_l}}{k_l!}}{\sum \limits_{k_3=0}^{\lfloor C/c_3 \rfloor} \sum \limits_{k_4=0}^{\lfloor (C - c_3k_3)/c_4 \rfloor} \ldots \sum \limits_{k_L=0}^{\lfloor(C - \sum \limits_{j=3}^{L-1}c_j k_j)/c_L \rfloor}\sum \limits_{t=3}^{L}k_t\left(\sum\limits_{i=3}^{L}k_i\right)!\prod \limits_{l=3}^{L} \frac{{|\mathcal{T}_l|}^{k_l}}{k_l!}},
\end{align}
where $c_j$ is the logical base gate implementation cost for $\mathcal{T}_j$, $k_l$ is the number of $\mathcal{T}_l$ gates within a particular sequence,~$|\mathcal{T}_l|$ is the number of gates within $\mathcal{T}_l$, and $L$ is the order of the Clifford hierarchy to include $Z$-rotation gates up to.

This calculation outputs values closely matching proportion results obtained using the sequence generation algorithm for random target gates, as shown in Figure~\ref{fig:proportions}. 
Figure~\ref{fig:proportions-t4} shows the summed proportions of all $\mathcal{T}_4$ gates among $\mathcal{T}_3 \cup \mathcal{T}_4$ gates over a variety of $\mathcal{T}_4$ cost values for sequences consisting of Set$_2$ logical base gates. Figure~\ref{fig:proportions-t5} shows the summed proportions of all~$\mathcal{T}_5$ gates among~$\mathcal{T}_3 \cup \mathcal{T}_4 \cup \mathcal{T}_5$ gates over a variety of~$\mathcal{T}_5$ cost values for sequences consisting of~Set$_3$ logical base gates. The other logical base gate costs are assigned values according to their distillation and implementation cost with a maximum logical base gate error of~$\mu = 10^{-15}$ as shown in Table~\ref{tab-costs}. These results suggest that increasing the logical base gate implementation cost of a set $\mathcal{T}_n$ drastically lowers the proportion of them found within the database of cost-optimal sequences. Thus they become less effective at reducing the average cost-optimal sequence costs since they are included within sequences less often. This is a simpler calculation compared to actually performing gate synthesis for many random target gates. The gate set proportions appears to give an indication for how useful the gate subset is among the rest of the base gates. We suspect there is potential that with some further research it could be used to help provide a quick approximation for how much the average synthesis cost reduces when including a base gate subset with specified cost values.

\section*{Discussion}
We investigated the cost of sequences produced by cost-optimal single-qubit gate synthesis using logical base gates from a combination of Clifford gates and $Z$-rotation gates from higher orders of the Clifford hierarchy. An algorithm, based on Dijkstra's algorithm, was used to generate a database of cost-optimal sequences from arbitrary single-qubit universal sets of logical base gates with individually assigned costs. As base gates, combinations of Clifford gates and $Z$-rotation gates from various orders of the Clifford hierarchy were used with two approaches of implementing them. The first uses a recursively applied $Z$-rotation catalyst circuit that utilises a temporary ancilla qubit, a small number of $T$ gates and a $Z$-rotation state to apply two $Z$-rotation gates of the same angle on two separate qubits while retaining the original $Z$-rotation state. We calculate average $T$-count costs for this approach using the following two methods: all output gates of the catalyst circuits are applied directly to target qubits; and each output gate is first applied to a $|+\rangle$ state to form an intermediate magic state, which is then consumed to implement the corresponding gate via a teleportation circuit. The second approach of implementing base gates is through magic state distillation and implementation circuits that can assign costs as the average number of raw magic states used to implement them in error-correction codes up to specified logical error rates. After assigning base gate costs using each method, gate synthesis was performed by finding nearest neighbours within the database of cost-optimal sequences in the Pauli vector space corresponding to combined gates of sequences. 

Using the $Z$-rotation catalyst approach with directly applied output gates to assign gate costs, we found that by including the higher order Clifford hierarchy $Z$-rotation gates along with the standard Clifford+$T$ set of base gates, there was a reduction in synthesis cost when compared to only using the Clifford+$T$ base gate set. The average cost-optimal sequence $T$-counts reduced by $34 \pm 3\%$, $42 \pm 2\%$, $49 \pm 2\%$, and $54 \pm 3\%$ for the accumulative inclusion of the fourth, fifth, sixth, and seventh orders respectively. When using the same approach but with all output gates being applied via intermediate magic states, the average cost-optimal sequence $T$-counts reduced by $29\pm 3\%$, $31\pm 3\%$, $31\pm 4\%$, and $31\pm 4\%$ for the accumulative inclusion of the fourth, fifth, sixth, and seventh orders respectively. Each average $T$-count calculated using the catalyst circuit approach assumes that every output gate of all recursive levels of the circuit are resourced such that no output gates are wasted. The procedure also assumes that there are sufficient numbers of ancilla qubits and $Z$-rotation catalyst states for smooth implementation of the gate sequences resulting from synthesis. A realistic employment of the approach would likely use a combination of direct application of output gates and the use of intermediate magic states. This is because direct application is cheaper with respect to $T$-count, however the intermediate magic states help make the implementation more flexible since they can be consumed at any time to implement the corresponding gate onto any target qubit. Nevertheless, these results show that there is potential for the average $T$-count to decrease by over 50\% when performing gate synthesis with higher order Clifford hierarchy $Z$-rotation base gates that are implemented using this approach, when compared to cost-optimal synthesis using only the Clifford+$T$ base gate set.

By instead using the magic state distillation approach with base gate costs assigned as the number of raw magic states, we found that including the fourth order $Z$-rotation gates from the Clifford hierarchy along with the standard Clifford+$T$ gate set decreased the average cost-optimal sequence costs by up to $30 \pm 2\%$. We observe a reduction of up to $33 \pm 2\%$ when additionally including the $Z$-rotation gates from the fifth order. No noticeable improvement is observed when additionally including higher order $Z$-rotation base gates up to the seventh order. Although these savings are not quite as large as what may be possible with the $Z$-rotation catalyst approach, the magic state distillation approach does not require an accessible collection of $Z$-rotation catalyst states to be stored throughout the computation. The implementation circuit for the distilled $Z$-rotation magic state does require the application of a double angled $Z$-rotation gate as a correction 50\% of the time. However, this correction gate can ideally be generated as it is required, so that every possible angled rotation does not need to be stored in advance. Also, the number of raw magic states is only a rough approximation for the actual resource costs of implementation. A precise calculation would be an extensive task that considers a variety of factors such as qubits count, circuit depth, magic state distillation cost and details of the error-correcting implementation.

We investigated the lack of further improvement found when including $Z$-rotation gates from higher than the fourth order of the Clifford hierarchy when using the direct magic state distillation approach and the $Z$-rotation catalyst circuit approach with output gates being applied via intermediate magic states. A model was developed that estimates the proportion of logical base gates within sequences approximating random target gates. This model assumes that each $Z$-rotation gate from orders three and above of the Clifford hierarchy have equal proportions when assigned equal cost values, that is, the gate operations have equal usefulness for approximating random target gates for the purposes of gate synthesis. The proportion estimations were shown to closely fit the data obtained using the sequence generation algorithm on random target gates. This suggests that the lack of observed cost reduction when using higher order logical base gates is due to there being far less numbers of them at their assigned costs within all cost-optimal sequences generated up to the chosen maximum sequence cost. Thus the frequency of the base gates being used for synthesis of random target gates is low, leading to a low level of influence over the average resource costs overall. The model provides a simple method, without needing to generate the full database of sequences, for estimating these gate proportions with each order of the Clifford hierarchy being assigned individual cost values.

\section*{Acknowledgements}
This work was supported by the University of Melbourne through the establishment of an IBM Q Network Hub at the University. CDH is supported by a research grant from the Laby Foundation. We would like to thank Earl Campbell and Kae Nemoto for valuable discussions. We would also like to thank the referee for suggesting the $Z$-rotation catalyst circuit approach for estimating gate costs.

\bibliographystyle{unsrtnat}

%\bibliography{references.bib}

\begin{thebibliography}{29}
\providecommand{\natexlab}[1]{#1}
\providecommand{\url}[1]{\texttt{#1}}
\expandafter\ifx\csname urlstyle\endcsname\relax
  \providecommand{\doi}[1]{doi: #1}\else
  \providecommand{\doi}{doi: \begingroup \urlstyle{rm}\Url}\fi

\bibitem[Bravyi and Kitaev(1998)]{bravyi1998quantum}
Sergey~B Bravyi and A~Yu Kitaev.
\newblock Quantum codes on a lattice with boundary.
\newblock \emph{arXiv preprint quant-ph/9811052}, 1998.

\bibitem[Dennis et~al.(2002)Dennis, Kitaev, Landahl, and
  Preskill]{dennis2002topological}
Eric Dennis, Alexei Kitaev, Andrew Landahl, and John Preskill.
\newblock Topological quantum memory.
\newblock \emph{Journal of Mathematical Physics}, 43\penalty0 (9):\penalty0
  4452--4505, 2002.
\newblock \doi{10.1063/1.1499754}.

\bibitem[Raussendorf et~al.(2007)Raussendorf, Harrington, and
  Goyal]{raussendorf2007topological}
Robert Raussendorf, Jim Harrington, and Kovid Goyal.
\newblock Topological fault-tolerance in cluster state quantum computation.
\newblock \emph{New Journal of Physics}, 9\penalty0 (6):\penalty0 199, 2007.
\newblock \doi{10.1088/1367-2630/9/6/199}.

\bibitem[Wang et~al.(2011)Wang, Fowler, and Hollenberg]{wang2011surface}
David~S Wang, Austin~G Fowler, and Lloyd~CL Hollenberg.
\newblock Surface code quantum computing with error rates over 1\%.
\newblock \emph{Physical Review A}, 83\penalty0 (2):\penalty0 020302, 2011.
\newblock \doi{10.1103/PhysRevA.83.020302}.

\bibitem[Eastin and Knill(2009)]{eastin2009restrictions}
Bryan Eastin and Emanuel Knill.
\newblock Restrictions on transversal encoded quantum gate sets.
\newblock \emph{Physical {R}eview {L}etters}, 102\penalty0 (11):\penalty0
  110502, 2009.
\newblock \doi{10.1103/PhysRevLett.102.110502}.

\bibitem[Zhou et~al.(2000)Zhou, Leung, and Chuang]{zhou2000methodology}
Xinlan Zhou, Debbie~W Leung, and Isaac~L Chuang.
\newblock Methodology for quantum logic gate construction.
\newblock \emph{Physical Review A}, 62\penalty0 (5):\penalty0 052316, 2000.
\newblock \doi{10.1103/PhysRevA.62.052316}.

\bibitem[Nielsen and Chuang(2010)]{nielsen_chuang_2010}
Michael~A. Nielsen and Isaac~L. Chuang.
\newblock \emph{The Solovay–Kitaev theorem}, page 617–624.
\newblock Cambridge University Press, 2010.
\newblock \doi{10.1017/CBO9780511976667.019}.

\bibitem[Kitaev et~al.(2002)Kitaev, Shen, and Vyalyi]{kitaev2002classical}
A~Yu Kitaev, AH~Shen, and MN~Vyalyi.
\newblock {Classical and Quantum Computation (Graduate Studies in Mathematics
  vol 47)(Providence, RI: American Mathematical Society)}.
\newblock 2002.
\newblock \doi{10.1090/GSM/047}.

\bibitem[Fowler(2011)]{fowler2011constructing}
Austin~G Fowler.
\newblock Constructing arbitrary {S}teane code single logical qubit
  fault-tolerant gates.
\newblock \emph{Quantum Information \& Computation}, 11\penalty0
  (9-10):\penalty0 867--873, 2011.

\bibitem[Dawson and Nielsen(2006)]{dawson2005solovay}
Christopher~M. Dawson and Michael~A. Nielsen.
\newblock The {S}olovay-{K}itaev algorithm.
\newblock \emph{Quantum Information \& Computation}, 6\penalty0 (1):\penalty0
  81–95, 2006.
\newblock ISSN 1533-7146.

\bibitem[Matsumoto and Amano(2008)]{matsumoto2008representation}
Ken Matsumoto and Kazuyuki Amano.
\newblock Representation of quantum circuits with {C}lifford and $\pi/8$ gates.
\newblock \emph{arXiv preprint arXiv:0806.3834}, 2008.

\bibitem[Bocharov and Svore(2012)]{bocharov2012resource}
Alex Bocharov and Krysta~M Svore.
\newblock Resource-optimal single-qubit quantum circuits.
\newblock \emph{Physical Review Letters}, 109\penalty0 (19):\penalty0 190501,
  2012.
\newblock \doi{10.1103/PhysRevLett.109.190501}.

\bibitem[Kliuchnikov et~al.(2013{\natexlab{a}})Kliuchnikov, Maslov, and
  Mosca]{kliuchnikov2012fast}
Vadym Kliuchnikov, Dmitri Maslov, and Michele Mosca.
\newblock Fast and efficient exact synthesis of single-qubit unitaries
  generated by {C}lifford and {$T$} gates.
\newblock \emph{Quantum {I}nformation \& {C}omputation}, 13\penalty0
  (7–8):\penalty0 607–630, 2013{\natexlab{a}}.
\newblock ISSN 1533-7146.

\bibitem[Kliuchnikov et~al.(2016)Kliuchnikov, Maslov, and
  Mosca]{kliuchnikov2016practical}
Vadym Kliuchnikov, Dmitri Maslov, and Michele Mosca.
\newblock Practical approximation of single-qubit unitaries by single-qubit
  quantum {C}lifford and {$T$} circuits.
\newblock \emph{IEEE Transactions on Computers}, 65\penalty0 (1):\penalty0
  161--172, 2016.
\newblock \doi{10.1109/TC.2015.2409842}.

\bibitem[Ross and Selinger(2016)]{ross2016optimal}
Neil~J Ross and Peter Selinger.
\newblock Optimal ancilla-free clifford+$t$ approximation of z-rotations.
\newblock \emph{Quantum Information \& Computation}, 16\penalty0
  (11-12):\penalty0 901--953, 2016.

\bibitem[Forest et~al.(2015)Forest, Gosset, Kliuchnikov, and
  McKinnon]{forest2015exact}
Simon Forest, David Gosset, Vadym Kliuchnikov, and David McKinnon.
\newblock Exact synthesis of single-qubit unitaries over {C}lifford-cyclotomic
  gate sets.
\newblock \emph{Journal of Mathematical Physics}, 56\penalty0 (8):\penalty0
  082201, 2015.
\newblock \doi{10.1063/1.4927100}.

\bibitem[Kliuchnikov et~al.(2015)Kliuchnikov, Bocharov, Roetteler, and
  Yard]{kliuchnikov2015framework}
Vadym Kliuchnikov, Alex Bocharov, Martin Roetteler, and Jon Yard.
\newblock A framework for approximating qubit unitaries.
\newblock \emph{arXiv preprint arXiv:1510.03888}, 2015.

\bibitem[Selinger(2015)]{selinger2012efficient}
Peter Selinger.
\newblock Efficient {C}lifford+{$T$} approximation of single-qubit operators.
\newblock \emph{Quantum Information \& Computation}, 15\penalty0
  (1–2):\penalty0 159–180, 2015.
\newblock ISSN 1533-7146.

\bibitem[Kliuchnikov et~al.(2013{\natexlab{b}})Kliuchnikov, Maslov, and
  Mosca]{kliuchnikov2013asymptotically}
Vadym Kliuchnikov, Dmitri Maslov, and Michele Mosca.
\newblock Asymptotically optimal approximation of single qubit unitaries by
  {C}lifford and {$T$} circuits using a constant number of ancillary qubits.
\newblock \emph{Physical {R}eview {L}etters}, 110\penalty0 (19):\penalty0
  190502, 2013{\natexlab{b}}.
\newblock \doi{10.1103/PhysRevLett.110.190502}.

\bibitem[Bocharov et~al.(2015)Bocharov, Roetteler, and
  Svore]{bocharov2015efficient}
Alex Bocharov, Martin Roetteler, and Krysta~M Svore.
\newblock Efficient synthesis of probabilistic quantum circuits with fallback.
\newblock \emph{Physical Review A}, 91\penalty0 (5):\penalty0 052317, 2015.
\newblock \doi{10.1103/PhysRevA.91.052317}.

\bibitem[Gottesman and Chuang(1999)]{gottesman1999demonstrating}
Daniel Gottesman and Isaac~L Chuang.
\newblock Demonstrating the viability of universal quantum computation using
  teleportation and single-qubit operations.
\newblock \emph{Nature}, 402\penalty0 (6760):\penalty0 390, 1999.
\newblock \doi{10.1038/46503}.

\bibitem[Gidney and Fowler(2019)]{gidney2019efficient}
Craig Gidney and Austin~G Fowler.
\newblock {Efficient magic state factories with a catalyzed $| CCZ\rangle $ to
  $2| T\rangle $ transformation}.
\newblock \emph{Quantum}, 3:\penalty0 135, 2019.
\newblock \doi{10.22331/Q-2019-04-30-135}.

\bibitem[Gidney(2018)]{gidney2018halving}
Craig Gidney.
\newblock Halving the cost of quantum addition.
\newblock \emph{Quantum}, 2:\penalty0 74, 2018.
\newblock \doi{10.22331/Q-2018-06-18-74}.

\bibitem[Campbell and O’Gorman(2016)]{campbell2016efficient}
Earl~T Campbell and Joe O’Gorman.
\newblock An efficient magic state approach to small angle rotations.
\newblock \emph{Quantum Science and Technology}, 1\penalty0 (1):\penalty0
  015007, 2016.
\newblock \doi{10.1088/2058-9565/1/1/015007}.

\bibitem[Campbell and Howard(2018)]{campbell2018magic}
Earl~T Campbell and Mark Howard.
\newblock Magic state parity-checker with pre-distilled components.
\newblock \emph{Quantum}, 2:\penalty0 56, 2018.
\newblock \doi{10.22331/Q-2018-03-14-56}.

\bibitem[Brown et~al.(2017)Brown, Bossomaier, and Barnett]{brown2017review}
Joshua~M Brown, Terry Bossomaier, and Lionel Barnett.
\newblock Review of data structures for computationally efficient
  nearest-neighbour entropy estimators for large systems with periodic boundary
  conditions.
\newblock \emph{Journal of {C}omputational {S}cience}, 23:\penalty0 109--117,
  2017.
\newblock \doi{10.1016/J.JOCS.2017.10.019}.

\bibitem[Uhlmann(1991)]{uhlmann1991satisfying}
Jeffrey~K. Uhlmann.
\newblock Satisfying general proximity/similarity queries with metric trees.
\newblock \emph{Information {P}rocessing {L}etters}, 40\penalty0 (4):\penalty0
  175--179, 1991.
\newblock \doi{10.1016/0020-0190(91)90074-r}.

\bibitem[Yianilos(1993)]{yianilos1993data}
Peter~N Yianilos.
\newblock Data structures and algorithms for nearest neighbor search in general
  metric spaces.
\newblock In \emph{Soda}, volume~93, pages 311--21, 1993.

\bibitem[Pham et~al.(2013)Pham, Van~Meter, and Horsman]{pham2013optimization}
Tien~Trung Pham, Rodney Van~Meter, and Clare Horsman.
\newblock Optimization of the {S}olovay-{K}itaev algorithm.
\newblock \emph{Physical Review A}, 87\penalty0 (5):\penalty0 052332, 2013.
\newblock \doi{10.1103/PhysRevA.87.052332}.

\end{thebibliography}

%\printbibliography
%\newpage{}
\section*{Appendix}
\appendix
\renewcommand\thefigure{\thesection.\arabic{figure}}
\setcounter{figure}{0} 
\renewcommand\thedefinition{\thesection.\arabic{definition}}
\setcounter{definition}{0} 
\renewcommand\thelemma{\thesection.\arabic{lemma}}
\setcounter{lemma}{0}
\renewcommand\thetheorem{\thesection.\arabic{theorem}}
\setcounter{theorem}{0}
\renewcommand\thecorollary{\thesection.\arabic{corollary}}
\setcounter{corollary}{0}

\section{Model for Gate Proportions}\label{sec:gate-proportions}
\begin{figure}
	\centering
	\includegraphics[width=0.55\textwidth]{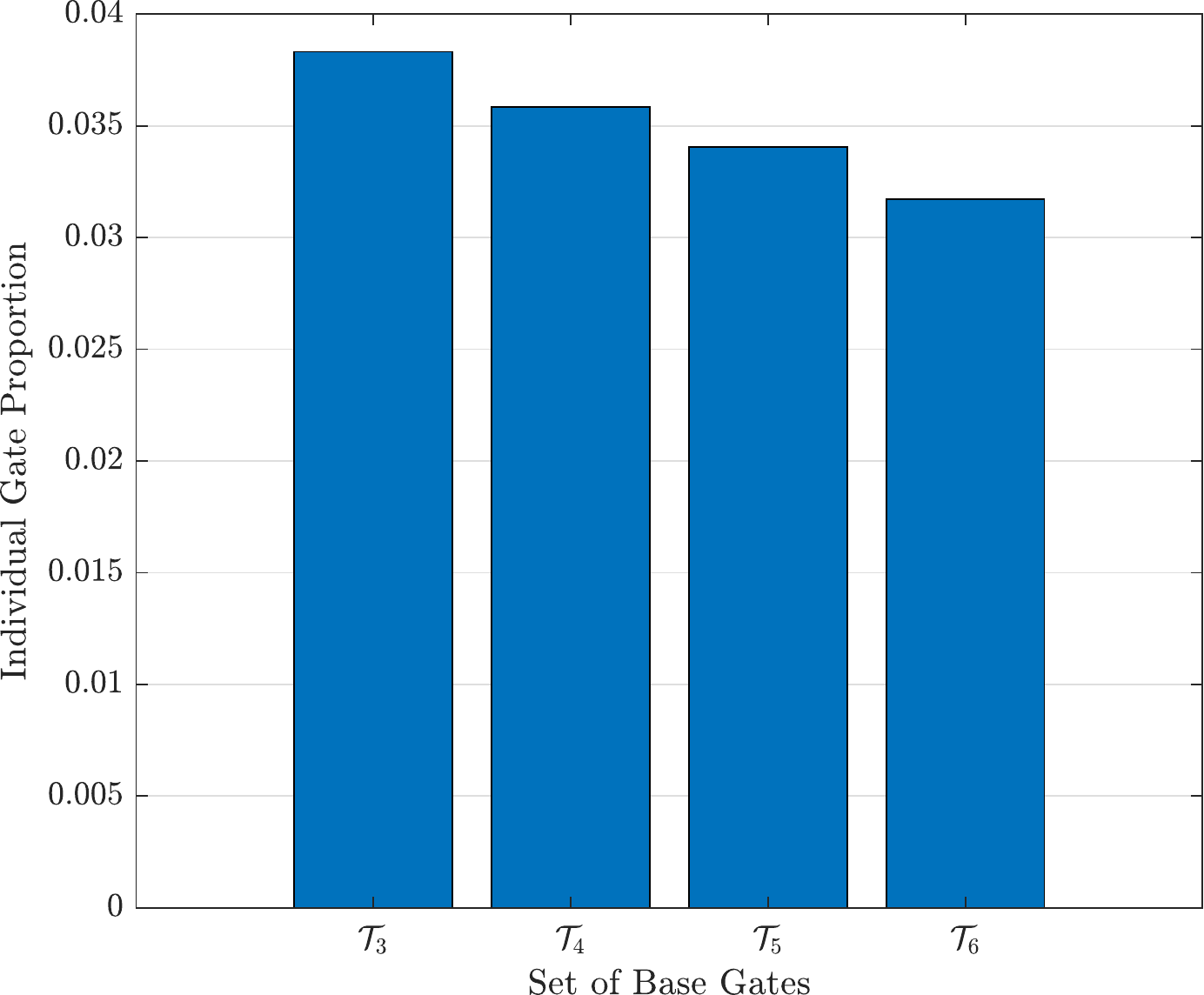}
	\caption{The proportions of individual logical base gates with equally assigned costs (synthesis logical level error $0.03$ using the trace distance). The number of gates within each set doubles for increasing order where $\mathcal{T}_3$ contains two gates (see Equation~\ref{eq:z-rotation-base-gate-sets}). This plot indicates that the logical base gates are almost equivalently useful in approximating random target gates using cost-optimal gate synthesis.}\label{fig:proportions-equal-cost}
\end{figure}

\begin{figure}
	\centering
	\textbf{Proportion of $\mathcal{T}_5$ gates for total sequence cost $C$}\par\medskip
	\includegraphics[width=0.55\textwidth]{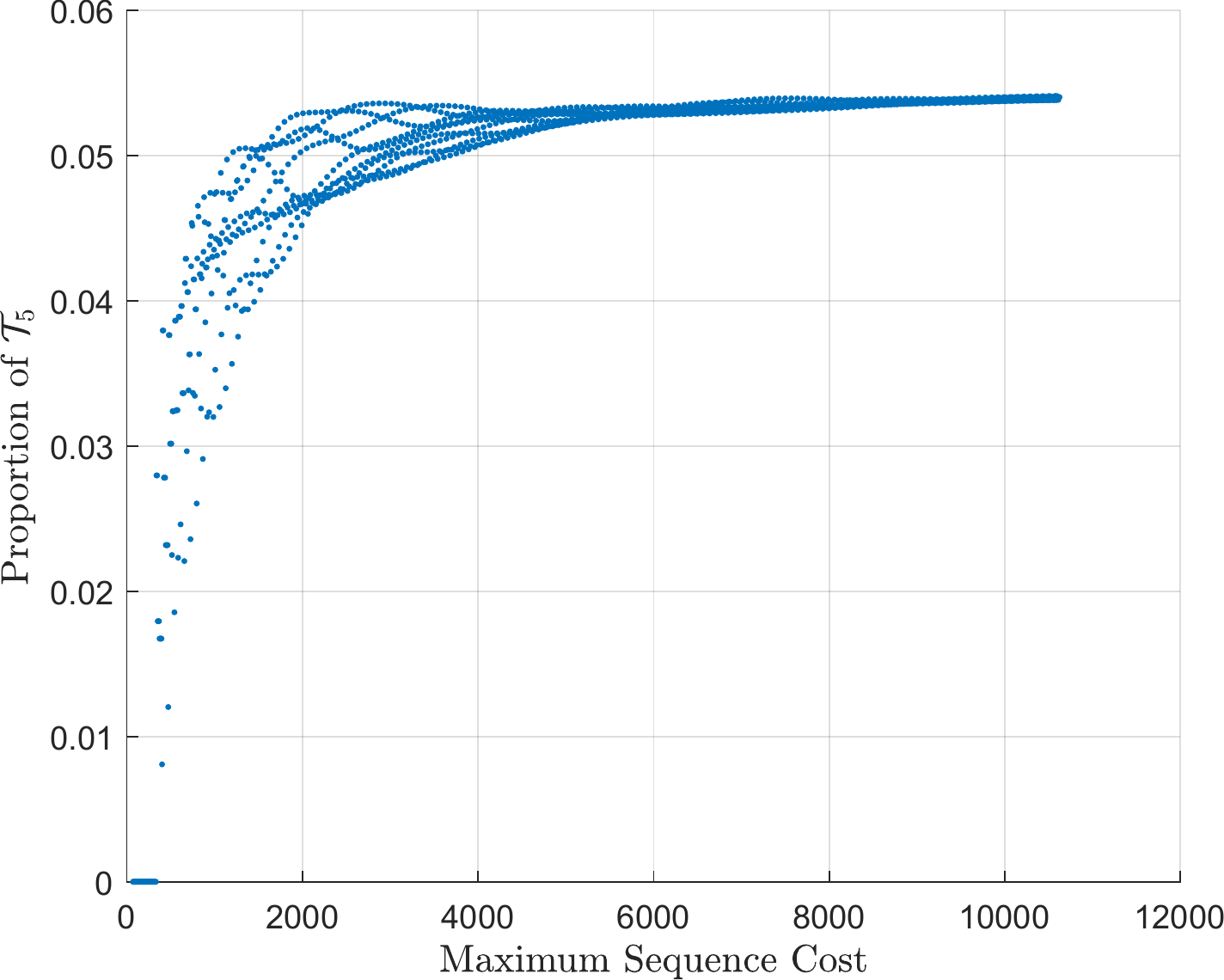}
	\caption{The proportion of $\mathcal{T}_5$ logical base gates among $\mathcal{T}_3 \cup \mathcal{T}_4 \cup \mathcal{T}_5$ gates calculated using the combinatorial model for all cost-optimal sequences below a maximum sequence cost that produce distinct combined gates.
	The logical base gate cost values are assigned according to Table~\ref{tab-costs} for a logical base gate error threshold of $\mu = 10^{-15}$ under the diamond norm. This plot shows that the proportion of $\mathcal{T}_5$ gates becomes approximately constant for sufficiently large maximum sequence costs. }\label{fig:t5-proportion-total-cost}
\end{figure}

Here we develop the theory for estimating the average proportion~$p_n$ of logical base gates among all~$\mathcal{T}_l$ gates (where~$l\geq 3$) with specified costs within cost-optimal sequences approximating random target gates synthesised to within an error threshold of~$\epsilon$. We begin by assuming that each logical base gate in $\mathcal{T}_3 \cup \mathcal{T}_4 \ldots \mathcal{T}_L$ for~$L\geq 3$ has equal proportions if they were to have equal costs, that is, the gate operations are equally effective for the purposes of gate synthesis. This can be justified by the data in Figure~\ref{fig:proportions-equal-cost}. The figure shows that when each logical base gate is given equal costs, the sequence generation algorithm generates a database of gate sequences with each gate having approximately the same proportions, where the proportions slowly decrease for increasing order. We do not expect these proportions to significantly change for larger sequence costs (or smaller synthesis error thresholds $\epsilon$) since the logical base gate proportions are approximately constant for sufficiently large maximum sequence costs. This can be seen in Fig.~\ref{fig:t5-proportion-total-cost} for the case of $\mathcal{T}_5$ logical base gates from within Set$_3$ generated by the sequence generation algorithm for random target gates.

Assume we have a database of cost-optimal gate sequences that have been generated up to a chosen maximum cost with individually assigned implementation costs for each set of logical base gates $\mathcal{T}_l$ where $l \geq 3$. We will calculate the proportion of $\mathcal{T}_n$ gates among all sequences within the database. For simplicity, let logical gates from any set~$\mathcal{T}_l$ for~$l\geq 3$ be called~$t$ gates. Using a unique canonical form~\cite{forest2015exact} for sequences consisting of the Clifford gates and combinations of~$\mathcal{T}_l$ gates, arbitrary gate sequences can be reduced to the form
\begin{equation}
    c . t_1 . H . t_2 . H \ldots t_{M} . c^\prime,\label{eq:canonical-form}
\end{equation}
where $c$ and $c^\prime$ are Clifford gates, $t_m$ is the $m^{\text{th}}$ positioned $t$ gate in the sequence, and $M$ is the $t$-count. For a particular sequence, let the number of $t$ gates from $\mathcal{T}_l$ be denoted by $k_l$. It follows that each sequence consisting of gates from up to order $L$ of the Clifford hierarchy satisfies (noting that~$c_0=c_1=0$)

\begin{equation}\label{eq:cost-constraint}
    \sum \limits_{l=3}^{L}c_l k_l \leq C,
\end{equation}
where $c_l$ is the cost assigned to logical gates from $\mathcal{T}_l$ and $C$ is the maximum cost of the database of gate sequences. It will be useful to denote the number of $t$ gates from order~$l$ to $L$ of the Clifford hierarchy as
\begin{equation}
    K_l := \sum \limits_{i=l}^{L}k_i,
\end{equation}
noting that $K_3$ is the $t$-count, $M$ that appears in Eq.~\ref{eq:canonical-form}.

The aim is to calculate the proportion of $\mathcal{T}_n$ gates among all gates in sequences within the database. We begin by counting the total number of possible sequences that can be formed given a set of $t$ gate counts~$\{k_l\}|_3^L$. Then the total number of possible sequences can be summed by iterating through every combination of possible sets $\{k_l\}|_3^L$ that satisfy Eq.~\ref{eq:cost-constraint} with their assigned base gate costs. Once this expression is determined, it can be extended to calculate the number of $\mathcal{T}_n$ gates and the total number of gates, which can then be used to calculate the proportions. For sequences of $t$-count~$K_3$, the number of permutations of $k_l$ gates within~$K_3$ gate locations is
\begin{equation}
    \mathrm{(\#Permutations(}k_l, K_3)):={K_3 \choose k_l} = \frac{K_3!}{(K_3-k_l)!k_l!}.
\end{equation}
Let $|\mathcal{T}_l|$ be the number of distinct $Z$-rotation gates within order $l$ of the Clifford hierarchy, for example, $|\mathcal{T}_3| = 2$ since $\mathcal{T}_3 = \{T, T^\dagger\}$ (up to global phase). Then for each permutation, there are $|\mathcal{T}_l|^{k_l}$ unique combinations of assigned $\mathcal{T}_l$ logical base gates within the permutation. Thus, the total number of configurations for $k_l$ number of gate locations with $|\mathcal{T}_l|$ variations in a sequence of $t$ gate count $K_3$ is 
\begin{equation}
    \gamma (k_l, |\mathcal{T}_l|, K_3) := {|\mathcal{T}_l|}^{k_l}\frac{K_3!}{(K_3-k_l)!k_l!}.
\end{equation}
After assigning gates to $k_l$ locations, there are $K_3 - k_l$ locations remaining within the sequence. The strategy from here is to iteratively count the total number of configurations from $l=3$ to $L$ by updating the number of remaining locations at each step, which now updates as  $K_{l+1} = K_l - k_l$. So for the second iteration, the number of configurations of $k_{l+1}$ gates with $|\mathcal{T}_{l+1}|$ variations within remaining locations $K_{l+1}$ of a given configuration from the assigned $k_l$ number of $\mathcal{T}_l$ gates is $\gamma (k_{l+1}, |\mathcal{T}_{l+1}|,K_{l+1})$, leading to a total of $\gamma(k_l, |\mathcal{T}_{l}|, K_l) \gamma(k_{l+1}, |\mathcal{T}_{l+1}|, K_{l+1})$ configurations for $k_l$ and $k_{l+1}$ numbers of $\mathcal{T}_l$ and $\mathcal{T}_{l+1}$ gates respectively in sequences of $t$-count $K_l$. Thus the total number of configurations for a set of $t$ gate counts~$\boldsymbol{k}=\{k_3, k_4, \ldots, k_L\}$ in sequences of $t$-count $K_3$ (containing $t$ gates up to order $L$ of the Clifford hierarchy) is
\begin{align}
    \Gamma(\boldsymbol{k}) &:= \prod \limits_{l=3}^{L}\gamma(k_l, |\mathcal{T}_l|, K_l)
    = \prod \limits_{l=3}^{L} {|\mathcal{T}_l|}^{k_l}\frac{K_l!}{(K_l-k_l)!k_l!} \\
    &= \frac{K_3! K_4! \ldots K_L!}{K_4! \ldots K_{L}! (K_L - k_L)!}\prod \limits_{l=3}^{L} \frac{{|\mathcal{T}_l|}^{k_l}}{k_l!}
    = K_3!\prod \limits_{l=3}^{L} \frac{{|\mathcal{T}_l|}^{k_l}}{k_l!} \\
    &= \left(\sum\limits_{i=3}^{L}k_i\right)!\prod \limits_{l=3}^{L} \frac{{|\mathcal{T}_l|}^{k_l}}{k_l!}.
\end{align}
To count the total number of sequences, we sum over all configurations for each assignment of $\boldsymbol{k}$ satisfying Equation~\ref{eq:cost-constraint}. We begin by determining the maximum allowable values for each $k_l$ with respect to already specified lower order $t$ gate counts~$\{k_j\}|_3^{l-1}$. The maximum possible value for $k_3$ is $\lfloor C/c_3 \rfloor$. Given a specified $k_3$, the maximum value for $k_{4}$ is~$\lfloor (C-c_3 k_3)/c_{4} \rfloor$. By continuing this pattern, given a set of~$t$ gate counts~$\{k_3, k_4, \ldots, k_{l-1}\}$, the maximum value for $k_l$ is 
\begin{equation}
    \text{max}(k_l) = \lfloor(C - \sum \limits_{j=3}^{l-1}c_j k_j)/c_i \rfloor.
\end{equation}
So now the total number of sequence configurations with logical base gate costs $\boldsymbol{c}$ and maximum sequence cost $C$ can be calculated as
\begin{align}
    \zeta(\boldsymbol{c}, C) &:= \sum \limits_{\{\boldsymbol{k} \;|\; \boldsymbol{c}\cdot\boldsymbol{k} \leq C\}} \Gamma(\boldsymbol{k}) \nonumber\\
    &= \sum \limits_{k_3=0}^{\lfloor C/c_3 \rfloor} \sum \limits_{k_4=0}^{\lfloor (C - c_3k_3)/c_4 \rfloor} \ldots \sum \limits_{k_L=0}^{\lfloor(C - \sum \limits_{j=3}^{L-1}c_j k_j)/c_L \rfloor}\left(\sum\limits_{i=3}^{L}k_i\right)!\prod \limits_{l=3}^{L} \frac{{|\mathcal{T}_l|}^{k_l}}{k_l!}.
\end{align}
Since the number of $\mathcal{T}_l$ logical gates within a particular sequence is $k_l$, the total number of $\mathcal{T}_l$ gates within all possible sequences below the maximum cost $C$ is calculated by multiplying $k_l$ to each term in the summation, the total number of gates can be calculated in a similar way. Thus, the proportion of $\mathcal{T}_n$ gates can be calculated as the weighted sum
\begin{align}
    p_n &= \frac{\sum \limits_{\{\boldsymbol{k} \;|\; \boldsymbol{c}\cdot\boldsymbol{k} \leq C\}} k_n\Gamma(\boldsymbol{k})}{\sum \limits_{\{\boldsymbol{k} \;|\; \boldsymbol{c}\cdot\boldsymbol{k} \leq C\}} \sum\limits_{t=3}^{L}k_t \Gamma(\boldsymbol{k})}\\
    &=\frac{\sum \limits_{k_3=0}^{\lfloor C/c_3 \rfloor} \sum \limits_{k_4=0}^{\lfloor (C - c_3k_3)/c_4 \rfloor} \ldots \sum \limits_{k_L=0}^{\lfloor(C - \sum \limits_{j=3}^{L-1}c_j k_j)/c_L \rfloor}k_n\left(\sum\limits_{i=3}^{L}k_i\right)!\prod \limits_{l=3}^{L} \frac{{|\mathcal{T}_l|}^{k_l}}{k_l!}}{\sum \limits_{k_3=0}^{\lfloor C/c_3 \rfloor} \sum \limits_{k_4=0}^{\lfloor (C - c_3k_3)/c_4 \rfloor} \ldots \sum \limits_{k_L=0}^{\lfloor(C - \sum \limits_{j=3}^{L-1}c_j k_j)/c_L \rfloor}\sum \limits_{t=3}^{L}k_t\left(\sum\limits_{i=3}^{L}k_i\right)!\prod \limits_{l=3}^{L} \frac{{|\mathcal{T}_l|}^{k_l}}{k_l!}}.
\end{align}
\newpage{}

\end{document}